\documentclass[12pt, letterpaper]{article}

\usepackage[margin=1in]{geometry}

\usepackage{setspace}
\usepackage{amsfonts,amssymb,amsbsy,amsxtra}

\usepackage{graphicx}
\usepackage{pdfpages}

\usepackage{booktabs}
\usepackage{longtable}
\usepackage{tabularx}
\usepackage{dcolumn}
\usepackage{multirow}
\usepackage{adjustbox}

\usepackage[dvipsnames,svgnames,table]{xcolor}
\definecolor{dark-red}{rgb}{0.75,0.10,0.10}
\definecolor{bluish}{rgb}{0.05,0.05,0.85}

\usepackage[unicode, naturalnames]{hyperref}
\hypersetup{
    colorlinks=true,
    linkcolor=blue,
    urlcolor=blue,
    citecolor=bluish,
    pdfstartview={XYZ null null 1.00},
    pdfpagemode=UseNone,
    pdftitle={blacklight}
}

\usepackage{natbib}
\usepackage{units}
\usepackage{nicefrac}

\usepackage[nameinlink, capitalize, noabbrev]{cleveref}

\usepackage{float}
\usepackage{placeins}

\usepackage[
    skip=0pt,
    labelfont=bf,
    font=small,
    textfont=small,
    figurename=Figure,
    justification=justified,
    singlelinecheck=false,
    labelsep=period
]{caption}
\usepackage{subcaption}

\usepackage{pdflscape}
\usepackage{rotating}

\usepackage{fancyhdr}

\usepackage{url}
\usepackage{xurl}

\usepackage{footmisc}
\usepackage[titletoc,title]{appendix}

\ExplSyntaxOn
\cs_if_exist:NF \expandableinput
{
  \cs_new:Npn \expandableinput #1
    { \use:c { @@input } { \file_full_name:n {#1} } }
}
\AddToHook{env/tabular/begin}
  { \cs_set_eq:NN \input \expandableinput }
\ExplSyntaxOff

\title{Exposed: Shedding \textit{Blacklight} on Online Privacy\thanks{Replication materials: \url{https://github.com/themains/private_blacklight}}}

\author{
    Lucas Shen\thanks{Agency for Science, Technology and Research. \href{mailto:lucas@lucasshen.com}{\texttt{lucas@lucasshen.com}}}
    \and
    Gaurav Sood\thanks{\href{mailto:gsood07@gmail.com}{\texttt{gsood07@gmail.com}}}
    \and
    Daniel Weitzel\thanks{Colorado State University. \href{mailto:daniel.weitzel@colostate.edu}{\texttt{daniel.weitzel@colostate.edu}}}
}
\date{\today}

\begin{document}
\maketitle
\onehalfspacing
\begin{abstract}

To what extent are users surveilled on the web, by what technologies, and by whom? We answer these questions by combining passively observed, anonymized browsing data of a large, representative sample of Americans with domain-level data on tracking from Blacklight. We find that nearly all users ($ > 99\%$) encounter at least one ad tracker or third-party cookie over the observation window. More invasive techniques like session recording, keylogging, and canvas fingerprinting are less widespread, but over half of the users visited a site employing at least one of these within the first 48 hours of the start of tracking. Linking trackers to their parent organizations reveals that a single organization, usually Google, can track over $50\%$ of web activity of more than half the users. Demographic differences in exposure are modest and often attenuate when we account for browsing volume. However, disparities by age and race remain, suggesting that what users browse, not just how much, shapes their surveillance risk.\\
    \noindent
	\textbf{Keywords:} Online Privacy, Online Safety, Digital Divide, KeyLogging, Tracking\\

        
\end{abstract}
\vspace{1cm}
\thispagestyle{empty}
\newpage
\setcounter{page}{1}
\newpage

\doublespacing
\section{Introduction}

The digital economy increasingly depends on personal data to mediate interactions between users, platforms, and advertisers. As individuals navigate the web, search for information, or engage with apps and services, their activity is routinely logged by a complex ecosystem of tracking technologies. These data flows enable large-scale personalization and behavioral advertising,  reshaping the online user experience.

From one perspective, the system has brought real benefits. For consumers, targeted advertising lowers search costs by highlighting products, services, or content that align with their preferences, potentially surfacing relevant options they might not otherwise encounter. For suppliers, especially smaller firms or new entrants, digital targeting offers a cost-effective way to reach relevant audiences without the inefficiencies of mass, untargeted advertising. This improved matching function can expand market reach for niche products and reduce customer acquisition costs. Data shows as much. Disabling cookies can reduce publisher revenue by over 50\% \citep{johnson2020consumer, ravichandran2019effect}, with the largest relative losses for small publishers and niche advertisers.

On the flip side, there are real costs to this system. The data that fuels personalization is often collected through opaque and increasingly invasive techniques, ranging from third-party cookies and fingerprinting to session recording and keylogging. These methods power a broader system of surveillance that can result in a wide array of harms. As \cite{citron2022privacy} argue, privacy violations can cause physical risks, e.g., stalking, economic losses, e.g., identity theft, psychological harms, e.g., anxiety or loss of trust, and reputational damage. They can also reinforce social inequality through discriminatory and exclusionary practices.

These concerns are magnified by the ease with which ostensibly anonymized data can be re-identified. Even datasets stripped of explicit identifiers can often be traced back to individuals using a small number of behavioral signals—such as search queries, media consumption patterns, or spatio-temporal traces from mobile devices \citep{achara2015unicity}. When such granular data becomes linkable across contexts, the potential for harm expands.

These risks are not merely hypothetical. In practice, they manifest in the form of predatory or discriminatory targeting. For instance, individuals facing financial hardship are disproportionately targeted with high-interest loans and other exploitative financial products \citep{christl2016networks}. As recent investigations have shown, users are steered toward more expensive options based on device type, e.g., Mac vs. PC, potentially reducing consumer surplus \citep{borgesius2020price, bujlow2015web, hannak2014measuring}. Relatedly, some work shows that advertisers and platforms engage in digital redlining, excluding certain users from seeing ads for housing, employment, or credit based on race, location, and other sensitive attributes \citep{angwin2016facebook}.

Despite widespread debate over the tradeoffs of online tracking, empirical evidence remains limited on where, how, and to whom these surveillance technologies are deployed.
Prior research has typically adopted a site-centric perspective, examining the prevalence of tracking technologies across websites \citep{Acar2013, Dambra2022, Englehardt2016, Iqbal2021, karaj2018whotracks, blacklight, niforatos2021prevalence, Nikiforakis2013, SanchezRola2018, SanchezRola2021, solomos2020clashtrackersmeasuringevolution, zheutlin2021data, zheutlin2022data}.
Yet this approach overlooks browsing behavior, and therefore considerably underestimates user-level exposure to tracking \citep{Dambra2022}.
How prevalent are advanced tracking techniques like fingerprinting or keylogging at the user level? Which types of users are more likely to encounter such techniques in their everyday browsing? Are certain populations, by virtue of the sites they visit, more exposed to surveillance than others?

This paper addresses these questions by combining two complementary data sources. We begin with passively collected, anonymized browsing data and sociodemographic profiles for a large, nationally representative sample of American adults, obtained from YouGov. These data provide granular insight into the websites people actually visit, enabling us to assess real-world exposure to tracking technologies rather than relying on stated privacy attitudes or a sample of highly visited sites. Each visited domain is then linked to privacy audit data from Blacklight, a tool developed by The Markup that scans websites for the presence of third-party cookies, device fingerprinting, session recording, keylogging, and redirect-based surveillance. This combined dataset enables us to assess the actual privacy risks that users face online and to quantify disparities in exposure across various demographics, including gender, race, education, and age. 

Our analysis offers three key contributions. First, we document the user-centric prevalence of sophisticated surveillance tools across the modern web. Second, we show how exposure varies across demographic groups, revealing new dimensions of digital inequality. Third, we provide a framework for measuring and monitoring privacy harms using passively collected behavioral data—a critical step toward evidence-based privacy policy and accountability.

\section{Research Design, Data, and Measures}
To quantify users’ exposure to online tracking, we combine two data sources: (1) a month-long, passively collected, anonymized dataset of domain-level web traffic from a nationally representative panel of 1,200 U.S. adults—covering over six million visits—and (2) domain-level audits from Blacklight, a real-time scanning tool developed by The Markup that detects seven types of tracking technologies, including more invasive techniques like session recording and canvas fingerprinting (\cref{sec:blacklight}).

We construct two complementary measures of user-level exposure (\cref{sec:measuring_exposure}). The first is cumulative exposure, the total number of tracker encounters during the observation window. The second is a rate-adjusted measure that normalizes by browsing volume, capturing the average number of trackers per visit. This distinction allows us to separate exposure due to time spent online from that driven by browsing choices. A very small number of panelists have no observed traffic during the study period and are excluded from the analyses. We assume these cases are missing completely at random. Similarly, not all visited domains return successful analyses from Blacklight, due to technical issues like temporary errors and redirects. These instances are excluded from the exposure computations and again assumed to be missing completely at random. We revisit these assumptions in \cref{sec:discussion}.

Beyond domain-level exposure, we assess how much of a user’s browsing trail is observable by the parent organization, e.g., Meta. To measure this surveillance capacity, we link third-party services to their parent firms and calculate the share of a user’s browsing history accessible to any one organization (\cref{sec:browsing_history}).

Lastly, we analyze demographic disparities in exposure (\cref{sec:demo_diff}), examining how age, race, gender, and education correlate with both the volume and rate of exposure.

\subsection{Browsing data}
Our browsing data comes from YouGov, which maintains a large panel of US adults and uses matched sampling to construct representative samples. This involves drawing a random population from a large synthetic representative sampling frame \citep{rivers2009}, who are then invited to take a survey. Non-respondents are replaced with similar individuals. Our study sample consists of 1,200 such American adults who have volunteered to install a passive metering software, RealityMine, on their device in lieu of rewards, which collects de-identified web browsing data over a one-month period in June 2022 \citep{bad_domains, data-dataverse, ss-adult}. This software logs visits to web domains with anonymized URLs (e.g., \textit{https://www.google.com/search?ANONYMIZED} or \textit{https://mail.google.com/mail/u/0/?ANONYMIZED}) and visit timestamps regardless of browser type or privacy settings. All participants gave informed consent and were fully aware of the data collection process, including passive web tracking, which they could opt out of at any time. Personal data such as passwords or secure form entries was excluded, with all data anonymized, including URLs as we described above (please see \cref{sm:yougov}).

\begin{table}[ht]
\captionsetup{justification=centering} 
\centering
\footnotesize
\setlength\tabcolsep{6 pt}
\setlength{\defaultaddspace}{0pt}
\def\sym#1{\ifmmode^{#1}\else\(^{#1}\)\fi}
\caption{Overview of data}
\label{tab:demo_summary}
\begin{adjustbox}{max width=\textwidth}
\begin{tabular}{@{\hspace{0\tabcolsep}}lrc@{\hspace{0\tabcolsep}}}
    \toprule\toprule
    \textbf{A. Sample size} & n & (\%)\\
    \midrule
    No. individuals & 1,132 & --- \\
    No. domains & 64,074 & ---\\
    No. visits & 6,297,382 & ---\\
    No. domains, Blacklight & 34,078 & (53.2\%) \\
    No. visits, Blacklight & 4,767,099 & (75.7\%) \\
    \midrule
    \textbf{B. Demographics} & n & (\%)\\
    \midrule
    Female & 635 & (52.9\%) \\
Male & 565 & (47.1\%) \\
White & 762 & (63.5\%) \\
Hispanic & 176 & (14.7\%) \\
Black & 152 & (12.7\%) \\
Other & 61 & (5.1\%) \\
Asian & 49 & (4.1\%) \\
High school diploma or below & 427 & (35.6\%) \\
Some College education & 350 & (29.2\%) \\
College Graduate & 272 & (22.7\%) \\
Postgraduate & 151 & (12.6\%) \\
$<$ 25 years old & 97 & (8.1\%) \\
25--34 years old & 222 & (18.5\%) \\
35--49 years old & 298 & (24.8\%) \\
50--64 years old & 301 & (25.1\%) \\
65+ years old & 282 & (23.5\%) \\
   \bottomrule
\end{tabular}
\end{adjustbox}
\caption*{\footnotesize Note:
    Percentages in Panel A represent the proportion of total domains or total visits covered by each tracking tool.
    Percentages in Panel B indicate the proportion of individuals in each demographic category.
}
\end{table}

Overall, our data of digital traces includes over 6 million web visits to over 64,000 unique domains from 1,134 individuals over a month (\cref{tab:demo_summary}). 65 individuals had no online activity on their device in the entire month, an additional individual had all visits without relevant metadata such as the URL, and two more had domains with no tracking data.

Our sample also includes individual-level demographics, summarized in Panel B of \cref{tab:demo_summary} such as gender, race (Black, Hispanic, White, Other), education level (high school diploma or below, some college education, college degree, postgraduate college degree), and age, which we bin into five groups: $<$ 25, 25--34, 35--49, 50--64, and 65+ years old. Our panel is representative of the US adult population, with the gender, race, education, age, and geography (five regions) closely resembling that of the same-year Current Population Survey \citep{bad_domains}.

\subsection{Measuring Tracking on Domains}
\label{sec:blacklight}
Blacklight is an on-demand privacy inspection tool that simulates a fresh user visiting a website and scans for seven types of stateful and stateless tracking methods. Blacklight identifies tracking through browser automation, network request monitoring, and behavioral script analysis. We submitted $64{,}074$ unique domains visited in our sample to Blacklight and obtained results for $34{,}078$ domains ($53.25\%$), covering $76\%$ of all visits in our dataset (\cref{tab:demo_summary}). Specifically, Blacklight detects these seven tracking methods (see \cref{sm:bl_tracking_mechanisms} for more details):
\begin{itemize}\itemsep0pt
    \footnotesize
    \item \textbf{Ad Trackers:} Detected via outgoing requests matched to DuckDuckGo’s ``Ad Motivated Tracking'' list.
    \item \textbf{Third-party Cookies:} Detected by analyzing `Set-Cookie` headers on requests to third-party services.
    \item \textbf{Facebook Pixel and Google Analytics:} Collect granular behavioral data for ad targeting and analytics.
    \item \textbf{Session Recording Scripts:} Detected based on script behavior and a known list of URLs for session replay services.
    \item \textbf{Keylogging:} Identified by typing known values into form fields and monitoring network activity for exfiltration of those exact keystrokes.
    \item \textbf{Canvas Fingerprinting:} Detected by inspecting `\texttt{<canvas>}` behavior and analyzing pixel-level script outputs.
\end{itemize}

Of these, session Recording, keylogging, and canvas fingerprinting are especially invasive \citep{Acar2014, karaj2018whotracks, blacklight, MS12, leakyforms}. These techniques also raise privacy risks beyond conventional tracking, as they bypass commonly proposed hygiene measures such as ad blockers and cookie deletion.

\subsection{Measuring Exposure to Tracking Methods}
\label{sec:measuring_exposure}

To quantify the extent of user-level exposure to online tracking, we link users' browsing data (\cref{sec:browsing_history}) with Blacklight scans for domain-level tracking (\cref{sec:blacklight}). Each individual $i$ has a set of site visits $\mathcal{V}_i$, where each visit $v$ corresponds to a timestamped instance of visiting a webpage from domain $d$. Let $d(v)$ denote the domain associated with visit $v$. $\vert \mathcal{V}_i \vert$ is the total number of visits for that individual in the month. We compute exposure to one of the tracking methods $s$ detected by Blacklight (\cref{sec:blacklight}) by aggregating tracker counts based on the domain of each visit (\cref{eq:cum_exposure}). To adjust for varying browsing intensity, we compute a rate-normalized exposure rate, normalizing cumulative exposure by the user's total number of visits (\cref{eq:exposure_rate}). 
\begin{align}
\text{Cumulative Exposure}_i^{(s)} &= 
    \sum_{v \in \mathcal{V}_i}
    \left\vert 
        trackers^{(s)}_{d(v)} 
    \right\vert,
\label{eq:cum_exposure} \\
\text{Exposure Rate}_i^{(s)} &= 
    \left(
        \frac{1}{\vert \mathcal{V}_i \vert}
        \cdot
        \text{Cumulative Exposure}_i^{(s)}
    \right) 
\label{eq:exposure_rate}
\end{align}

These measures approximate the cumulative volume and rate of behavioral data collected on an individual, reflecting the size of their digital footprint.
We use these metrics to examine the extent of privacy exposure online and disparities across demographic groups, leveraging self-reported characteristics collected alongside the browsing data (\cref{sec:demo_diff}).
In subsequent analyses, we use both measures to examine the extent of individual privacy exposure and its variation across demographic subgroups (\cref{sec:demo_diff}).

\subsection{Measuring Tracking by Organizations: Browsing History}\label{sec:browsing_history}

\begin{equation}\label{eq:org_per_user}
\left| \text{Organizations}_i \right| = 
\left| 
\bigcup_{v \in \mathcal{V}_i} O_{iv} 
\right|
\end{equation}
\begin{equation}\label{eq:ind_org_share_visits}
\text{Tracking share}_{ij} = 
\frac{
    \sum_{v \in \mathcal{V}_i}
    \mathbf{1}
    (j \in O_{iv})
}
{
    \vert \mathcal{V}_i \vert
}
\end{equation}

To measure the breadth and depth of tracking by organizations, we link domain-level metadata from the Blacklight analyses, which identifies the third-party domains (e.g., \textit{connect.facebook.net}) embedded on the private domains, to parent organizations (e.g., \textit{Facebook, Inc.}) using the DuckDuckGo Tracker Radar data (\url{https://github.com/duckduckgo/tracker-radar}).
The Tracker Radar maps over 38,000 third-party domains to over 19,000 distinct organizations. We then link these parent organizations ($O$) to the visit-level data ($\mathcal{V}$) via the detected third-party domains. This allows us to quantify: (i) the number of distinct organizations tracking each user (\cref{eq:org_per_user}) and (ii) how much of a user's browsing activity is visible to any organization $j$ (\cref{eq:ind_org_share_visits}). Organizations owning multiple third-party domains on the same private domain are counted only once.%
\footnote{
    We also compute organizations' shares weighted by dwelling time ($t$):
    \begin{equation}\label{eq:ind_org_share_duration}
        \text{Tracking share}_{ij}^{\text{(dur)}} = 
        \frac{
            \sum_{v \in \mathcal{V}_i} 
            \mathbf{1}
            (j \in O_{iv})
            \cdot
            t_{iv}
        }
        {
            \sum_{v \in \mathcal{V}_i} 
            t_{iv}
        },
    \end{equation}
    as an alternative measure of \cref{eq:ind_org_share_visits}, and reach similar findings (\cref{sm:time_weighted_org_tracking}).
}

\subsection{Demographic Differences}\label{sec:demo_diff}
To estimate disparities in online tracking, we model cumulative exposure and exposure rate as a function of a person's demographics. Specifically, 
\begin{equation}\label{eq:demo-diff}
     \text{y}_i =
     \alpha
     \, + \beta_1   \text{women}_i
     \, + \beta_2^k \text{race}_i
     \, + \beta_3^k \text{education}_i
     \, + \beta_4^k \text{age group}_i^k
     \, + \varepsilon_i,
\end{equation}
where the outcome measure is individual $i$'s exposure to each of the seven tracking methods from Blacklight. All models are estimated using ordinary least squares with Huber-White robust standard errors. Demographic covariates include gender (woman; ref: man), race/ethnicity (African American, Asian, Hispanic, Other; ref: White), education (some college, college degree, postgraduate; ref: high school or less), and age group (25–34, 35–49, 50–64, 65+; ref: 18–24). Since all demographic predictors are represented as indicator variables, their coefficients can be compared directly.

\section{Results}
\label{sec:results}

The results section is structured as follows. First, we report the prevalence and speed of exposure to the seven tracking technologies. Second, we examine how exposure varies by demographics. Third, we quantify the extent to which a single tracking organization can observe a user's online activity. Finally, we examine demographic differences in the depth of tracking by organizations.

\subsection{Exposure to Different Kinds of Tracking}
\label{sec:prevalence_tracking}

\begin{table}[ht]
\centering
\small
\setlength\tabcolsep{6 pt}
\setlength{\defaultaddspace}{0pt}
\def\sym#1{\ifmmode^{#1}\else\(^{#1}\)\fi}
\caption{Summary of cumulative exposure}
\label{tab:cumulative-exposure}
\begin{adjustbox}{max width=\textwidth}
\begin{tabular}{@{\hspace{0\tabcolsep}}lrrrrrrrrr@{\hspace{0\tabcolsep}}}
    \toprule\toprule
    &\multicolumn{7}{c}{Cumulative exposure}&\multicolumn{2}{c}{Percentage encountering}\\
    \cmidrule(lr){2-8}\cmidrule(l){9-10}
    & Mean & Std. dev. & Min. & 25p & Median & 75p & Max. & At least 1 & At least 10\\
    &\multicolumn{1}{r}{(1)}         &\multicolumn{1}{r}{(2)}         &\multicolumn{1}{r}{(3)}         &\multicolumn{1}{r}{(4)}         &\multicolumn{1}{r}{(5)}         &\multicolumn{1}{r}{(6)}         &\multicolumn{1}{r}{(7)}&\multicolumn{1}{r}{(8)}&\multicolumn{1}{r}{(9)}\\
    \midrule
    Ad Trackers & 27,407 & 48,279 & 0 & 2,620 & 9,738 & 29,240 & 517,968 & 99.6\% & 99.1\% \\
Third-Party Cookies & 32,325 & 55,184 & 0 & 3,133 & 11,757 & 35,647 & 700,142 & 99.4\% & 99.1\% \\
Facebook Pixel & 383 & 657 & 0 & 40 & 147 & 463 & 5,808 & 94.7\% & 87.3\% \\
Google Analytics & 35 & 104 & 0 & 0 & 8 & 29 & 1,619 & 72.4\% & 46.5\% \\
Session Recording & 155 & 353 & 0 & 10 & 54 & 165 & 5,788 & 89.7\% & 76.0\% \\
Keylogging & 309 & 935 & 0 & 4 & 26 & 148 & 10,315 & 84.9\% & 65.9\% \\
Canvas Fingerprinting & 320 & 697 & 0 & 18 & 84 & 288 & 7,643 & 91.7\% & 81.0\% \\
   \bottomrule
\end{tabular}
\end{adjustbox}
\caption*{\footnotesize Note:
    Cumulative exposure to trackers is defined in \cref{eq:cum_exposure}.
    Columns (8)--(9) report the percentage of people encountering at least one and at least ten trackers within the month.
}
\end{table}

\begin{table}[ht]
\centering
\small
\setlength\tabcolsep{12 pt}
\setlength{\defaultaddspace}{0pt}
\def\sym#1{\ifmmode^{#1}\else\(^{#1}\)\fi}
\caption{Summary of exposure rate}
\label{tab:exposure-rate}
\begin{adjustbox}{max width=\textwidth}
\begin{tabular}{@{\hspace{0\tabcolsep}}lrrrrrrr@{\hspace{0\tabcolsep}}}
    \toprule\toprule
    & Mean & Std. dev. & Min. & 25p & Median & 75p & Max.\\
    &\multicolumn{1}{r}{(1)}         &\multicolumn{1}{r}{(2)}         &\multicolumn{1}{r}{(3)}         &\multicolumn{1}{r}{(4)}         &\multicolumn{1}{r}{(5)}         &\multicolumn{1}{r}{(6)}         &\multicolumn{1}{r}{(7)}\\
    \midrule
    Ad Trackers & 4.98 & 3.64 & 0.00 & 2.66 & 3.99 & 6.28 & 31.41 \\
Third-Party Cookies & 6.12 & 5.05 & 0.00 & 3.26 & 4.83 & 7.36 & 53.42 \\
Facebook Pixel & 0.08 & 0.09 & 0.00 & 0.03 & 0.06 & 0.11 & 1.00 \\
Google Analytics & 0.01 & 0.04 & 0.00 & 0.00 & 0.00 & 0.01 & 0.99 \\
Session Recording & 0.03 & 0.05 & 0.00 & 0.01 & 0.02 & 0.04 & 0.59 \\
Keylogging & 0.04 & 0.07 & 0.00 & 0.00 & 0.01 & 0.04 & 0.58 \\
Canvas Fingerprinting & 0.06 & 0.09 & 0.00 & 0.01 & 0.04 & 0.07 & 1.00 \\
   \bottomrule
\end{tabular}
\end{adjustbox}
\caption*{\footnotesize Note:
    Exposure rates to trackers are defined in \cref{eq:exposure_rate}.
}
\end{table}

Tracking is near universal, with ad trackers and third-party cookies the most common methods. During the month-long observation period, $99.1\%$ of users encountered more than ten ad trackers or third-party cookies  (see \cref{tab:cumulative-exposure}). On average, users encountered $27{,}407$ ad trackers ($\hat{\sigma} = 48{,}279$) and $32{,}325$ third-party cookies ($\hat{\sigma} = 55{,}184$). The corresponding medians---$9{,}738$ and $11{,}757$ (\cref{tab:cumulative-exposure})---suggest heavily right-skewed distributions. Normalizing by the number of visits dramatically reduces the skew. Users are exposed to, on average, $5$ ad trackers ($\hat{\sigma} = 3.6$) and $6.1$ third-party cookies ($\hat{\sigma} = 5.1$) per visit (\cref{tab:exposure-rate}) with medians of $4$ and $4.8$ respectively. A tighter spread and lower skew suggest that most of the variation in total exposure is driven by differences in how much users browse. 

More invasive tracking methods---session recording, keylogging, and canvas fingerprinting---can be found on nearly $9\%$ of the domains but are encountered less frequently. Users encounter session recording on $3\%$ of visits ($\hat{\sigma} = 0.05$), keylogging scripts on $4\%$ of visits ($\hat{\sigma} = 0.07$), and fingerprinting scripts on $6\%$ of visits ($\hat{\sigma} = 0.09$) (\cref{tab:exposure-rate}). This suggests users are likelier to browse domains without invasive tracking. Despite the low rates, the cumulative exposure is non-trivial. For instance, $91.7\%$ of users encountered canvas fingerprinting at least once, and over $65\%$ encountered all three at least ten times (\cref{tab:cumulative-exposure}).

\begin{figure}[ht]
\subcaptionsetup{justification=centering}
  \centering
  \begin{subfigure}{\textwidth}
    \centering
    \includegraphics[width=0.55\textwidth]{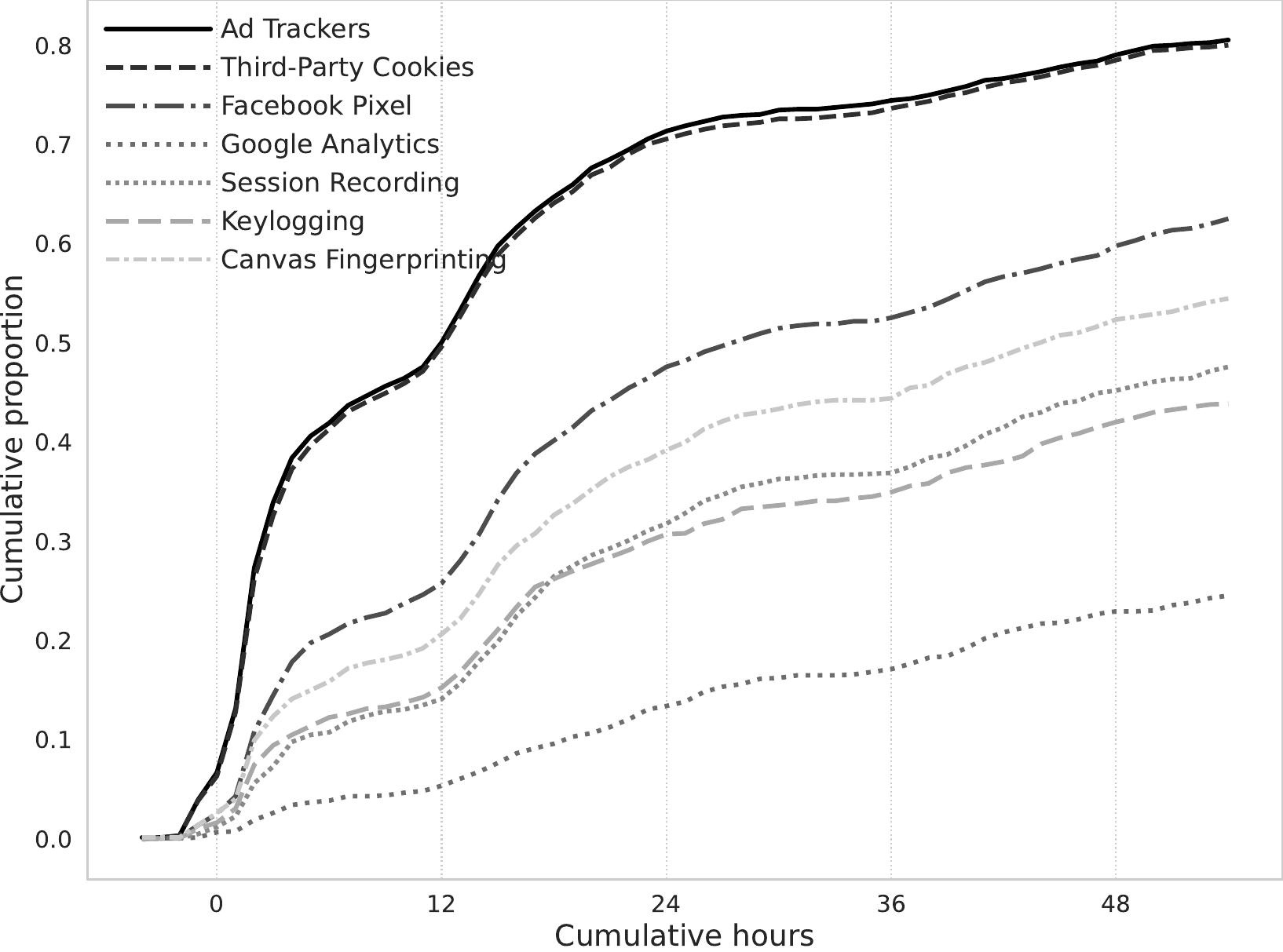}
    \caption{}
  \end{subfigure}
  \begin{subfigure}[b]{\textwidth}
    \centering
    \scriptsize
    \begin{tabular}{@{\hspace{0\tabcolsep}}lccccc@{\hspace{0\tabcolsep}}}
      \toprule
      &0h& 12h & 24h & 36h & 48h\\
      \input{tables/cum_exposure_by_hour}
      \bottomrule
    \end{tabular}
    \caption{}
  \end{subfigure}
\caption{
    \footnotesize
   The proportion of users who had encountered a particular tracker by a particular time. We start measuring at 6 PM on 31 May (due to time zones) when at least 50 users have logged browsing activity.
   The table reports the cumulative proportions at the specified hours.
}
\label{fig:cum_exposure_by_hour}
\end{figure}

Because tracking is pervasive, exposure is rapid. Using browsing timestamps, we identify when each user first encountered each tracking method. Half of the users encounter an ad tracker or a third-party cookie within the first $12$ hours of the start of measurement (see \cref{fig:cum_exposure_by_hour}). By 48 hours, nearly $80\%$ have encountered at least one tracker or cookie. Even the more intrusive techniques---session recording, keylogging, and canvas fingerprinting---reach nearly half the users within $48$ hours.

\subsection{Demographic Differences in Exposure to Tracking Methods}
\label{sec:demo_diff_mechanisms}

\begin{table}[!ht]
\centering
\small
\setlength\tabcolsep{3 pt}
\setlength{\defaultaddspace}{0pt}
\def\sym#1{\ifmmode^{#1}\else\(^{#1}\)\fi}
\caption{Demographic differences in cumulative exposure}
\label{tab:demo_differences_cum_exposure}
\begingroup
\centering
\begin{adjustbox}{max width=\textwidth}
\begin{tabular}{@{\hspace{0\tabcolsep}}l*{8}{D{.}{.}{-1}}@{\hspace{0\tabcolsep}}}
      \toprule
      &\multicolumn{7}{c}{Tracking mechanisms}\\
      \cmidrule(lr){2-8}
      &\multicolumn{1}{c}{Ads}&\multicolumn{1}{c}{Cookies}&\multicolumn{1}{c}{FB Pixel}&\multicolumn{1}{c}{GA}&\multicolumn{1}{c}{Keyloggers}&\multicolumn{1}{c}{Session rec}&\multicolumn{1}{c}{Canvas FP}&\multicolumn{1}{c}{Max share}\\
      &\multicolumn{1}{c}{(1)}&\multicolumn{1}{c}{(2)}&\multicolumn{1}{c}{(3)}&\multicolumn{1}{c}{(4)}&\multicolumn{1}{c}{(5)}&\multicolumn{1}{c}{(6)}&\multicolumn{1}{c}{(7)}&\multicolumn{1}{c}{(8)}\\
      \midrule 
      Woman                   & -35.4         & -30.5         & -38.1         & 2.1           & -0.96         & -2.8          & 92.9$\sym{**}$   & -5.4$\sym{**}$\\   
                              & (27.5)        & (31.4)        & (38.8)        & (6.0)         & (55.3)        & (20.7)        & (39.1)        & (2.7)\\   
      Race: African American  & -18.6         & -27.6         & -1.8          & -7.0          & -32.8         & -3.3          & -39.0         & -2.5\\   
                              & (41.4)        & (45.0)        & (69.5)        & (7.7)         & (83.7)        & (26.3)        & (63.0)        & (4.2)\\   
      Race: Asian             & 11.3          & 34.5          & 21.8          & 39.6          & -141.1$\sym{*}$  & -58.7$\sym{**}$  & 16.8          & 13.2\\   
                              & (69.9)        & (83.0)        & (88.3)        & (33.8)        & (76.5)        & (25.5)        & (80.4)        & (9.2)\\   
      Race: Hispanic          & -40.6         & -41.7         & -76.9$\sym{**}$  & -17.2$\sym{***}$ & -53.8         & -19.7         & -24.2         & -2.5\\   
                              & (30.6)        & (35.3)        & (37.3)        & (5.3)         & (72.6)        & (24.9)        & (51.0)        & (3.8)\\   
      Race: Other             & -13.9         & -2.9          & -10.8         & -12.3         & -137.8$\sym{**}$ & -33.0         & 191.8         & -4.4\\   
                              & (57.9)        & (75.2)        & (90.7)        & (7.8)         & (67.5)        & (27.8)        & (146.2)       & (4.8)\\   
      Educ: Some college      & 9.9           & 27.9          & 46.8          & 11.2          & -17.1         & 4.8           & 32.7          & 4.1\\   
                              & (29.6)        & (35.2)        & (44.2)        & (7.1)         & (65.1)        & (19.9)        & (43.8)        & (3.0)\\   
      Educ: College           & 126.3$\sym{***}$ & 160.9$\sym{***}$ & 76.8          & 15.4$\sym{**}$   & 169.4$\sym{*}$   & 87.8$\sym{**}$   & 87.2$\sym{*}$    & 13.0$\sym{***}$\\   
                              & (43.4)        & (49.9)        & (48.4)        & (7.0)         & (87.4)        & (35.9)        & (52.7)        & (3.8)\\   
      Educ: Postgraduate      & 89.9$\sym{*}$    & 87.6$\sym{*}$    & 173.0$\sym{**}$  & 13.0          & 6.2           & 68.2$\sym{*}$    & 160.9$\sym{*}$   & 11.1$\sym{**}$\\   
                              & (52.1)        & (52.4)        & (88.1)        & (13.3)        & (79.6)        & (35.4)        & (94.2)        & (4.9)\\   
      Age: 25--34             & 20.2          & 22.4          & 31.1          & -8.4          & -95.8         & 0.27          & 35.6          & 2.9\\   
                              & (25.1)        & (31.7)        & (54.1)        & (13.6)        & (116.5)       & (32.2)        & (51.4)        & (5.4)\\   
      Age: 35--49             & 99.0$\sym{***}$  & 97.9$\sym{***}$  & 75.7          & 6.3           & 94.6          & 37.4          & 84.2$\sym{**}$   & 2.5\\   
                              & (30.6)        & (34.9)        & (50.9)        & (13.2)        & (127.2)       & (32.5)        & (42.3)        & (4.8)\\   
      Age: 50--64             & 185.9$\sym{***}$ & 217.8$\sym{***}$ & 175.8$\sym{***}$ & -4.2          & 146.8         & 75.5$\sym{**}$   & 152.5$\sym{***}$ & 7.5\\   
                              & (39.3)        & (46.0)        & (57.3)        & (12.0)        & (133.8)       & (36.9)        & (45.7)        & (5.1)\\   
      Age: 65+                & 309.3$\sym{***}$ & 351.7$\sym{***}$ & 320.3$\sym{***}$ & 3.3           & 287.9$\sym{**}$  & 136.0$\sym{***}$ & 358.6$\sym{***}$ & 13.7$\sym{***}$\\   
                              & (37.5)        & (44.7)        & (65.0)        & (12.0)        & (132.1)       & (36.7)        & (59.6)        & (4.9)\\   
      Constant                & 110.0$\sym{***}$ & 125.5$\sym{***}$ & 217.4$\sym{***}$ & 28.3$\sym{***}$  & 188.1$\sym{*}$   & 73.8$\sym{**}$   & 69.4$\sym{*}$    & 21.5$\sym{***}$\\   
                              & (29.2)        & (33.9)        & (50.3)        & (10.4)        & (110.9)       & (30.0)        & (41.6)        & (4.5)\\   
       \\
      Dependent variable mean & 274.1         & 323.3         & 383.3         & 35.1          & 309.1         & 155.4         & 319.8         & 30.1\\  
      R$^2$                   & 0.07          & 0.07          & 0.04          & 0.02          & 0.03          & 0.04          & 0.05          & 0.04\\  
      Observations & \multicolumn{1}{c}{\text{1,134}} & \multicolumn{1}{c}{\text{1,134}} & \multicolumn{1}{c}{\text{1,134}} & \multicolumn{1}{c}{\text{1,134}} & \multicolumn{1}{c}{\text{1,134}} & \multicolumn{1}{c}{\text{1,134}} & \multicolumn{1}{c}{\text{1,134}} & \multicolumn{1}{c}{\text{1,134}}\\
      \bottomrule
   \end{tabular}
\end{adjustbox}
\par\endgroup
\caption*{
    \footnotesize \textit{Note:}
    Each column reports coefficients from estimating \cref{eq:demo-diff},
    where the outcome is the cumulative exposure (\cref{eq:cum_exposure}) to the seven tracking mechanisms and the number of visits tracked by the top organization in column (8), defined as the tracker organization associated with the highest share of a user's total web visits (\cref{sec:browsing_history}).
    Ad trackers (Ads) and third-party cookies (columns 1–2), and the max share of visits tracked (column 8) are scaled by a factor of $\nicefrac{1}{100}$, such that a coefficient of 1 corresponds to 100 tracking instances.
    Please see \cref{fig:coefplot_demo_differences_cumulative} for an alternative visualization of the estimates.
    Significance levels: $^{*}$ 0.1 $^{**}$ 0.05 $^{***}$ 0.01.
}
\end{table}
\begin{table}[!ht]
\centering
\small
\setlength\tabcolsep{0 pt}
\setlength{\defaultaddspace}{0pt}
\def\sym#1{\ifmmode^{#1}\else\(^{#1}\)\fi}
\caption{Demographic differences in exposure rate}
\label{tab:demo_differences_exposure_rate}
\begingroup
\centering
\begin{adjustbox}{max width=\textwidth}
\begin{tabular}{@{\hspace{0\tabcolsep}}l*{8}{D{.}{.}{-1}}@{\hspace{0\tabcolsep}}}
      \toprule
      &\multicolumn{7}{c}{Tracking mechanisms}\\
      \cmidrule(lr){2-8}      
      &\multicolumn{1}{c}{Ads}&\multicolumn{1}{c}{Cookies}&\multicolumn{1}{c}{FB Pixel}&\multicolumn{1}{c}{GA}&\multicolumn{1}{c}{Keyloggers}&\multicolumn{1}{c}{Session rec}&\multicolumn{1}{c}{Canvas FP}&\multicolumn{1}{c}{Max share}\\
      &\multicolumn{1}{c}{(1)}&\multicolumn{1}{c}{(2)}&\multicolumn{1}{c}{(3)}&\multicolumn{1}{c}{(4)}&\multicolumn{1}{c}{(5)}&\multicolumn{1}{c}{(6)}&\multicolumn{1}{c}{(7)}&\multicolumn{1}{c}{(8)}\\
      \midrule 
      Woman                   & -0.203        & -0.101        & 0.002         & -0.0009       & 0.000  & 0.004                  & 0.011$\sym{**}$  & -0.017$\sym{*}$\\   
                              & (0.211)       & (0.291)       & (0.005)       & (0.002)       & (0.004)                & (0.003)                & (0.005)       & (0.010)\\   
      Race: African American  & -0.035        & -0.453        & 0.004         & 0.0003        & 0.004                  & 0.009                  & -0.0007       & -0.018\\   
                              & (0.339)       & (0.430)       & (0.010)       & (0.003)       & (0.007)                & (0.006)                & (0.008)       & (0.015)\\   
      Race: Asian             & -1.20$\sym{***}$ & -1.49$\sym{***}$ & -0.020$\sym{**}$ & -0.002        & -0.018$\sym{***}$         & -0.013$\sym{***}$         & -0.014$\sym{*}$  & 0.006\\   
                              & (0.299)       & (0.436)       & (0.009)       & (0.003)       & (0.005)                & (0.004)                & (0.007)       & (0.030)\\   
      Race: Hispanic          & 0.088         & 0.040         & 0.0006        & -0.001        & 0.001                  & 0.000  & 0.003         & -0.0002\\   
                              & (0.322)       & (0.452)       & (0.008)       & (0.003)       & (0.007)                & (0.004)                & (0.007)       & (0.015)\\   
      Race: Other             & -0.279        & -0.093        & -0.008        & -0.004$\sym{**}$ & -0.009                 & 0.002                  & 0.012         & -0.010\\   
                              & (0.435)       & (0.672)       & (0.008)       & (0.002)       & (0.008)                & (0.006)                & (0.011)       & (0.021)\\   
      Educ: Some college      & 0.192         & 0.188         & -0.002        & -0.0002       & 0.000   & 0.003                  & 0.008         & 0.023$\sym{*}$\\   
                              & (0.265)       & (0.362)       & (0.007)       & (0.003)       & (0.005)                & (0.004)                & (0.007)       & (0.012)\\   
      Educ: College           & 0.490$\sym{*}$   & 0.761$\sym{*}$   & -0.010        & -0.003        & 0.006                  & 0.002                  & 0.001         & 0.039$\sym{***}$\\   
                              & (0.294)       & (0.421)       & (0.007)       & (0.003)       & (0.006)                & (0.004)                & (0.007)       & (0.013)\\   
      Educ: Postgraduate      & 0.245         & 0.265         & -0.007        & -0.005        & -0.0002                & 0.004                  & 0.017$\sym{*}$   & 0.054$\sym{***}$\\   
                              & (0.315)       & (0.440)       & (0.008)       & (0.003)       & (0.007)                & (0.005)                & (0.010)       & (0.016)\\   
      Age: 25--34             & 0.375         & 0.412         & -0.013        & -0.004        & 0.0008                 & 0.002                  & 0.004         & -0.035\\   
                              & (0.279)       & (0.427)       & (0.014)       & (0.005)       & (0.008)                & (0.006)                & (0.009)       & (0.022)\\   
      Age: 35--49             & 1.82$\sym{***}$  & 1.95$\sym{***}$  & 0.006         & 0.003         & 0.018$\sym{**}$           & 0.012$\sym{*}$            & 0.016         & -0.032\\   
                              & (0.315)       & (0.480)       & (0.014)       & (0.006)       & (0.008)                & (0.006)                & (0.010)       & (0.020)\\   
      Age: 50--64             & 1.92$\sym{***}$  & 2.12$\sym{***}$  & 0.0004        & -0.003        & 0.026$\sym{***}$          & 0.015$\sym{**}$           & 0.009         & -0.071$\sym{***}$\\   
                              & (0.312)       & (0.431)       & (0.013)       & (0.005)       & (0.008)                & (0.007)                & (0.009)       & (0.019)\\   
      Age: 65+                & 2.81$\sym{***}$  & 3.07$\sym{***}$  & 0.006         & -0.005        & 0.035$\sym{***}$          & 0.014$\sym{**}$           & 0.033$\sym{***}$ & -0.066$\sym{***}$\\   
                              & (0.318)       & (0.449)       & (0.013)       & (0.005)       & (0.009)                & (0.006)                & (0.009)       & (0.019)\\   
      Constant                & 3.27$\sym{***}$  & 4.20$\sym{***}$  & 0.085$\sym{***}$ & 0.014$\sym{***}$ & 0.021$\sym{***}$          & 0.020$\sym{***}$          & 0.036$\sym{***}$ & 0.587$\sym{***}$\\   
                              & (0.264)       & (0.400)       & (0.014)       & (0.005)       & (0.007)                & (0.006)                & (0.009)       & (0.019)\\   
       \\
      Dependent variable mean & 5.0           & 6.1           & 0.08          & 0.010         & 0.04                   & 0.04                   & 0.06          & 0.55\\  
      R$^2$                   & 0.07          & 0.05          & 0.01          & 0.01          & 0.03                   & 0.02                   & 0.03          & 0.03\\  
      Observations & \multicolumn{1}{c}{\text{1,134}} & \multicolumn{1}{c}{\text{1,134}} & \multicolumn{1}{c}{\text{1,134}} & \multicolumn{1}{c}{\text{1,134}} & \multicolumn{1}{c}{\text{1,134}} & \multicolumn{1}{c}{\text{1,134}} & \multicolumn{1}{c}{\text{1,134}} & \multicolumn{1}{c}{\text{1,134}}\\
      \bottomrule
   \end{tabular}
\end{adjustbox}
\par\endgroup
\caption*{
    \footnotesize \textit{Note:}
    Each column reports coefficients from estimating \cref{eq:demo-diff},
    where the outcome is the exposure rate (\cref{eq:exposure_rate}) to the seven tracking mechanisms and the share of users' visits tracked by the top organization in column (8), defined as the tracker organization associated with the highest share of a user's total web visits (\cref{sec:browsing_history}).
    Please see \cref{fig:coefplot_demo_differences_rate} for an alternative visualization of the estimates.
    Significance levels: $^{*}$ 0.1 $^{**}$ 0.05 $^{***}$ 0.01.
}
\end{table}

\cref{tab:demo_differences_cum_exposure} and \cref{tab:demo_differences_exposure_rate} (columns (1)--(7)) report regression estimates of demographic differences in cumulative exposure and exposure rate for different tracking methods. 

Controlling for other demographic factors, gender is not a strong predictor of net exposure to tracking---except, women encounter canvas fingerprinting significantly more than men ($\hat{\beta} = 93$, $\widehat{\mathrm{SE}} = 39.1$, $p < .05$) (see \cref{tab:demo_differences_cum_exposure}). Racial differences are also limited: Asians are less exposed to keylogging and session recording, those categorized as `Others' are less exposed to keylogging, and Hispanic users are tracked less frequently by Facebook Pixel and Google Analytics.

In contrast, differences by education and age are more pronounced. College-educated users encounter more trackers than those with a high school diploma or less. For instance, they encounter $16{,}090$ more third-party cookies ($p < .01$) and $88$ more session recorders ($\widehat{\mathrm{SE}}$ = $35.9$, $p < .05$). Users with a postgraduate degree show similar patterns to those with a college degree. Age also plays a large role: older users are most exposed, with those 65 and above encountering significantly more trackers across all tracking methods, except for Google Analytics.

\begin{figure}[ht]
  \centering
  \includegraphics[width=.55\textwidth]{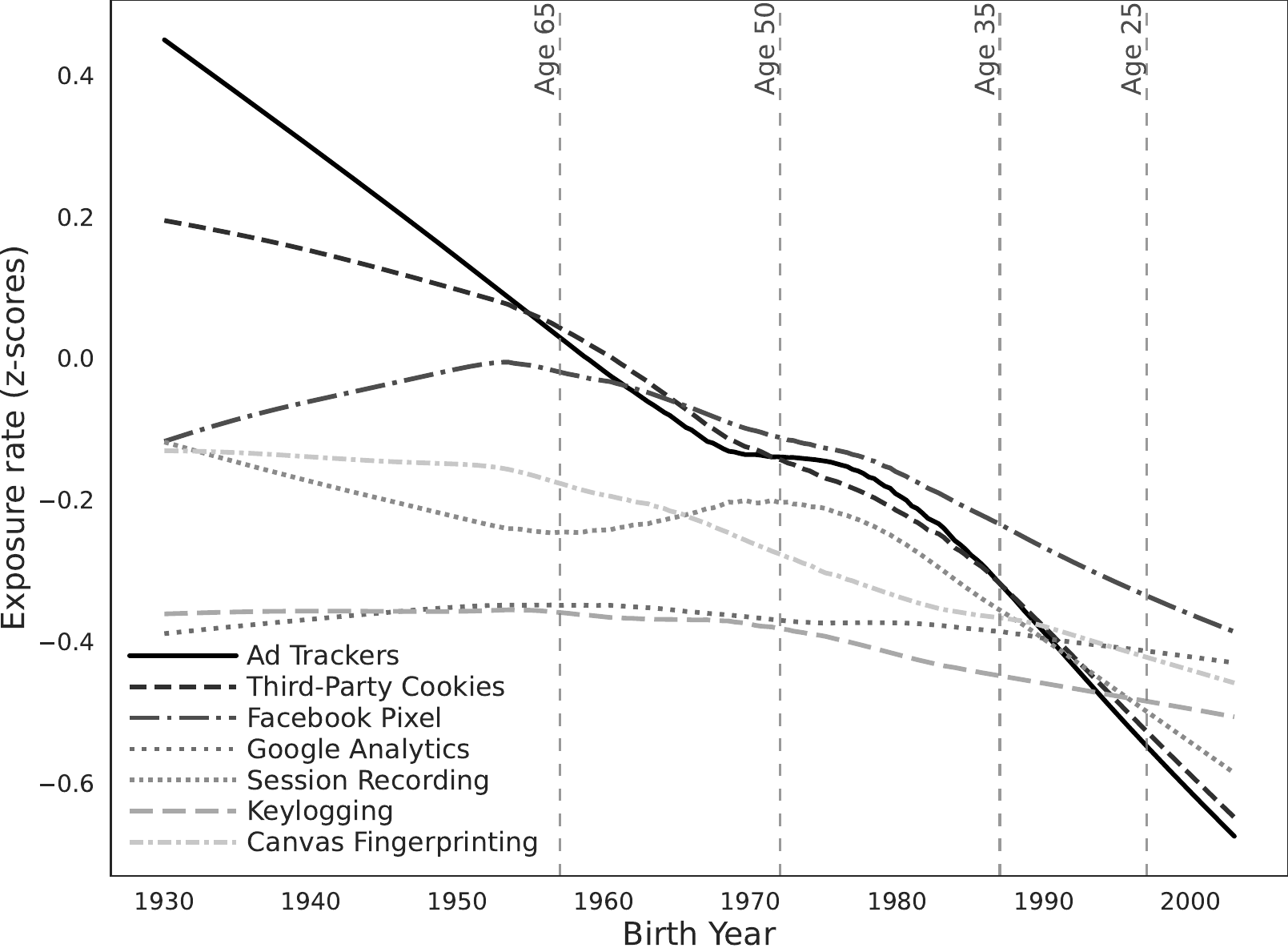}
  \caption{
    \footnotesize
    Exposure rate by birth year.
    Lines represent LOWESS-smoothed standardized rates (z-scores) of the exposure rates by the seven tracking methods.
    Values are winsorized at the 95th percentile.
    Vertical dashed lines correspond to the age groups. 
  }
  \label{fig:lowess_age_bl}
\end{figure}

Adjusting for browsing volume suggests that some demographic differences in tracking exposure reflect how much people browse, not which sites they visit (see \cref{tab:demo_differences_exposure_rate}). For example, the large gaps by education mostly vanish after normalization, suggesting that more educated users are online more often—not browsing more heavily tracked sites.

Some differences, however, remain. The gender gap in canvas fingerprinting remains: women encounter one additional fingerprinting script per $100$ visits ($\widehat{\mathrm{SE}} = 0.005$, $p < .05$). Age gradients in exposure also remain. Older users—especially those 65 and above—continue to experience higher exposure rates to ad trackers, third-party cookies, session recording, keylogging, and canvas fingerprinting (see \cref{fig:lowess_age_bl}).%
\footnote{
    \label{fn:bonferroni}
    Many demographic differences in exposure rates are significant even after correcting for multiple comparisons. Applying a Bonferroni correction for the $12$ demographic predictors tested ($p < .00416$), all coefficients with unadjusted $p < .01$ in \cref{tab:demo_differences_exposure_rate} remain significant.
}

Some differences sharpen after normalization.  Asian users, who had lower cumulative exposure only to session recording and keylogging, now show lower exposure rates across nearly every method but Google Analytics. This suggests that, once online activity is held constant, they tend to visit less heavily tracked sites.

Taken together, these results help pinpoint the sources of demographic gaps in tracking. Some reflect how often people go online; others reflect where they go. However, it is important to note that demographics explain little on their own: across all models, they account for less than 8\% of the variation in exposure  \crefrange{tab:demo_differences_cum_exposure}{tab:demo_differences_exposure_rate}), pointing to the dominant role of individual browsing habits.

\subsection{Tracking by Organizations}
\label{sec:tracker_organizations}

\begin{figure}
    \centering
    \begin{subfigure}[t]{0.495\textwidth}
        \centering
        \includegraphics[width=\textwidth]{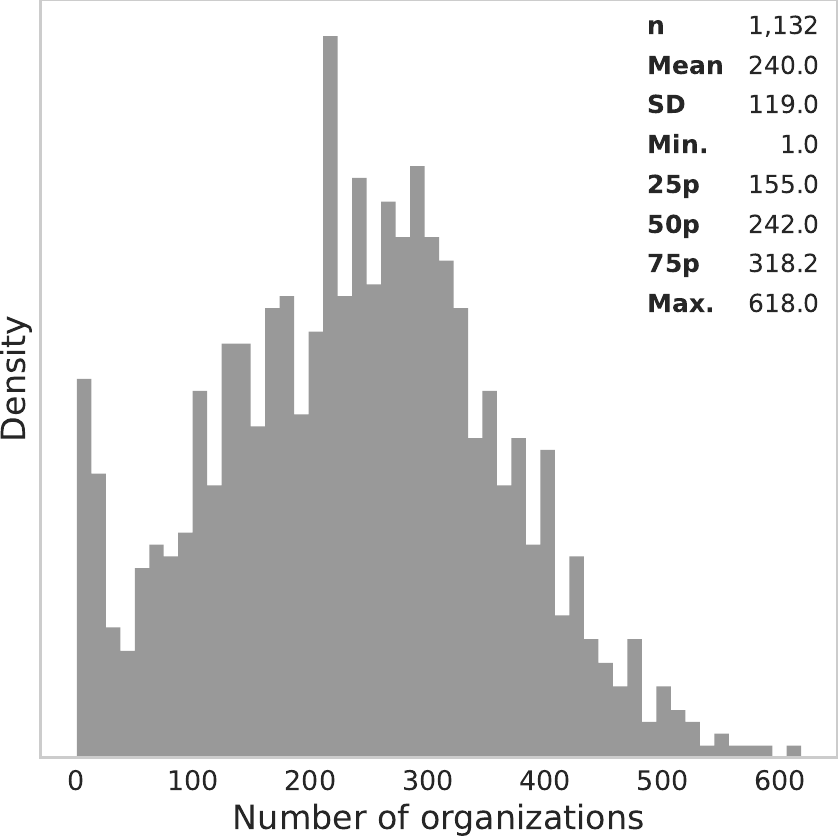}
        \caption{Number of organizations}
        \label{fig:dist_org_per_user_summtable}
    \end{subfigure}
    \hfill
    \begin{subfigure}[t]{0.495\textwidth}
        \centering
        \includegraphics[width=\linewidth]{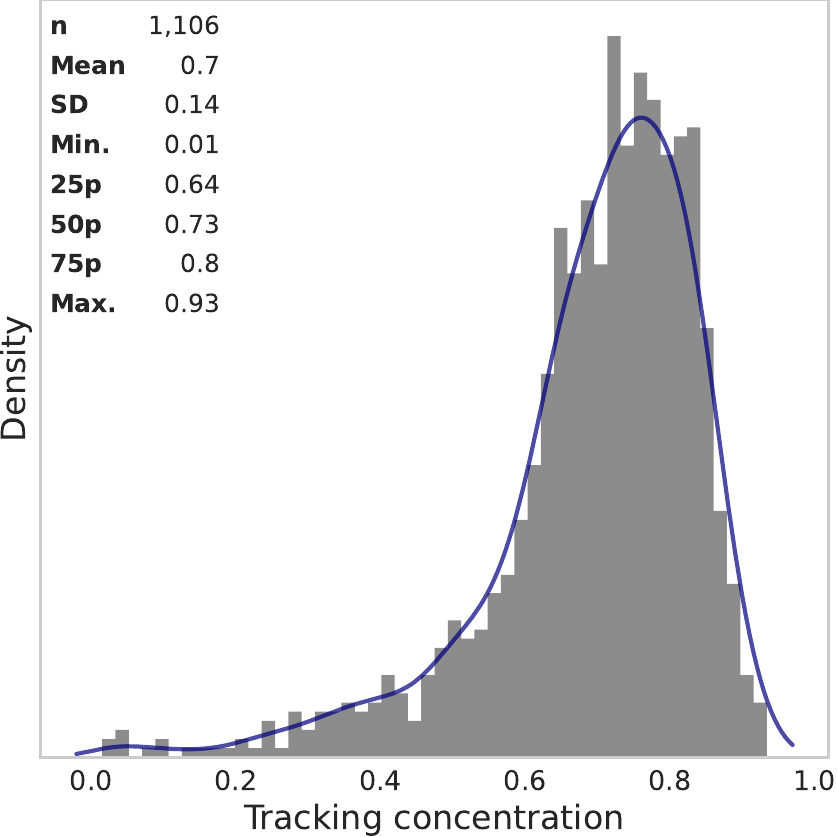}
        \caption{Organization tracking concentration}
        \label{fig:dist_tracking_concentration_per_user_summtable}
    \end{subfigure}    
    
    \vspace{0.5cm} 
    
    \begin{subfigure}[t]{0.495\textwidth}
        \centering
        \includegraphics[width=\linewidth]{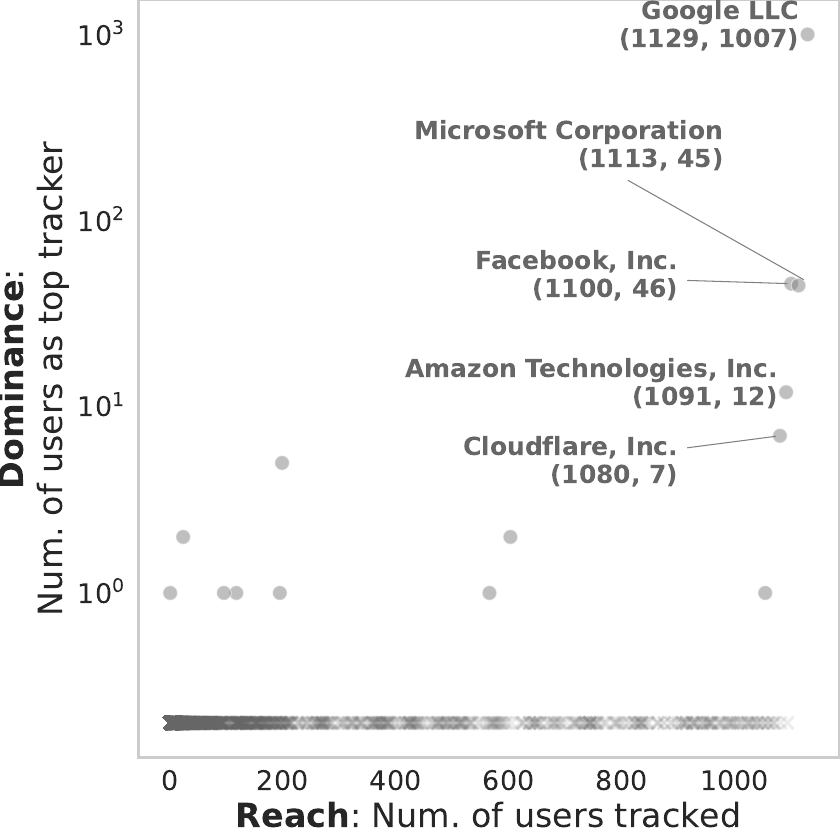}
        \caption{Dominance vs. Reach}
        \label{fig:top_trackers_reach_dominance_annotated_top_orgs}
    \end{subfigure}
    \hfill
    \begin{subfigure}[t]{0.495\textwidth}
        \centering
        \includegraphics[width=\textwidth]{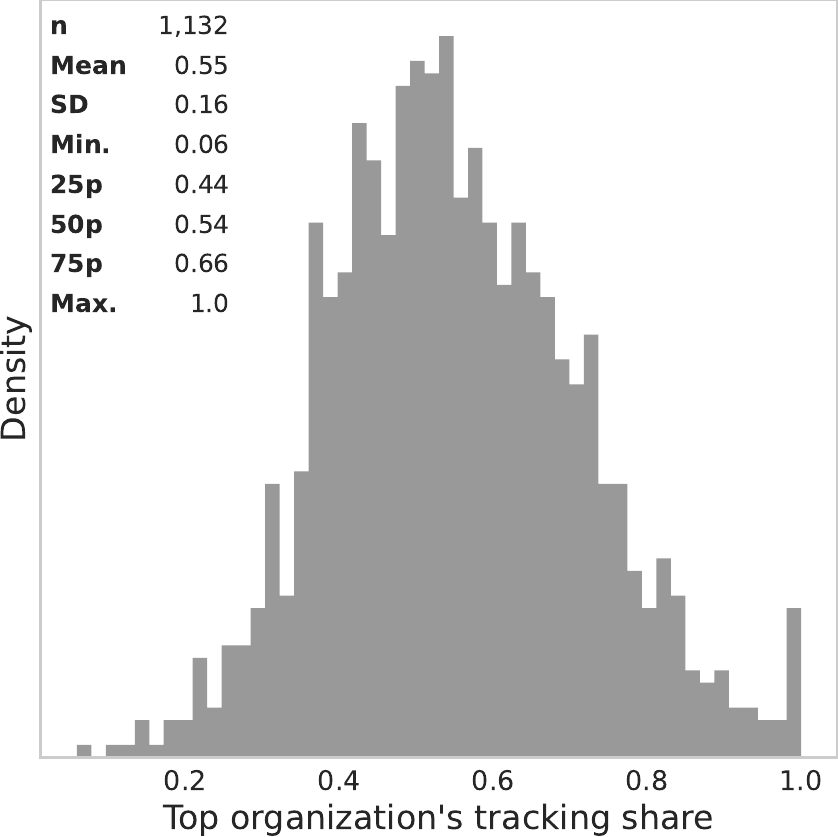}
        \caption{User-level browsing history tracked by the top organization}
        \label{fig:dist_maxshare_hist_summtable}
    \end{subfigure}

    \caption{
        \footnotesize
        The share of browsing history tracked by parent organizations.
        Panel (a) reports the number of organizations tracking each user’s browsing history.
        Panel (b) reports the concentration of users' browsing history exposure across organizations (Gini coefficients, for those tracked by $\geq 10$ organizations).
        Panel (c) plots each organization's dominance---the number of users for whom it tracked the largest share of browsing history---against its Reach, the total number of users it tracked.
        The parentheses report the corresponding numbers.
        Panel (d) reports the largest share of each user's browsing history tracked by a single organization.
    }
    \label{fig:tracking_by_organizations}
\end{figure}

Mapping third-party services to parent organizations, we assess both the number of organizations tracking each user (\cref{eq:org_per_user}) and the share of users’ browsing histories tracked by each organization (\cref{eq:ind_org_share_visits}). 

\cref{fig:dist_org_per_user_summtable} shows that users are typically tracked by $155$ to $318$ organizations, with a median of $242$. Despite this breadth, exposure is highly concentrated. \cref{fig:dist_tracking_concentration_per_user_summtable} shows that for users tracked by at least ten organizations, exposure is dominated by a handful of organizations, with the median Gini coefficient of $0.73$.

\cref{fig:top_trackers_reach_dominance_annotated_top_orgs} plots organizations' tracking \textit{dominance}---the number of users for whom it has the largest share of browsing history---against tracking \textit{reach}---the number of users it tracks at least once, highlighting organizations with near-ubiquitous presence. Google towers over all in both reach and dominance, being the top organization for $99.6\%$ of the sample. Other prominent organizations are Microsoft, Facebook, Amazon, and Cloudflare.%
\footnote{
    Likewise, tracking exposure is highly concentrated among a handful of domains (\cref{sm:top_tracking_domains}), with many sites embedding multiple types of tracking technologies. Financial and e-commerce platforms are particularly prominent in contributing to the tracking via session recording and keylogging, while other big tech and social media companies, such as Microsoft and TikTok, are prominent in canvas fingerprinting.
}

\cref{fig:dist_maxshare_hist_summtable} shows the distribution of the maximum share of browsing history of a user tracked by an organization. On average, $55\%$ of a user's browsing history is tracked by a single organization ($\hat{\sigma} = 0.16$). The median user has similar exposure, with $54\%$ of their browsing history tracked by any single organization. At the 75th percentile, the top organization's share is $66\%$. Defining organizations' tracking share using the time spent online (\cref{eq:ind_org_share_duration}) yields similar measures (\cref{sm:time_weighted_org_tracking}).

\subsection{Demographics Differences in Tracking by Organizations}
\label{sec:demo_diff_organizations}

Lastly, we consider how the share of a user's browsing activity visible to the single most dominant tracking organization varies by demographics.

Whereas \cref{sec:demo_diff_mechanisms} examines demographic differences exposure to the seven tracking technologies detected by Blacklight, here we examine demographic differences in (i) the cumulative share of total visits observed by the top organization (column (8), \cref{tab:demo_differences_cum_exposure}) and (ii) the rate-normalized proportion of total visits observed by the top organization (column (8), \cref{tab:demo_differences_exposure_rate}).

Women have a slightly lower depth of exposure than men, while those with a college degree or postgraduate education have a greater depth of exposure compared to those with a high school diploma or below. These differences hold even when normalized by total visits (see column (8) of \cref{tab:demo_differences_exposure_rate}). Women have a $1.7$ percentage point lower maximum share of visits ($\widehat{\mathrm{SE}}$ = $1.0\%$, $p < .1$), while college-educated and postgraduate users have $3.9$ (SE = $1.3\%$, $p < .01$) and $5.4$ ($\widehat{\mathrm{SE}}$ = $1.6\%$, $p < .01$) percentage point higher shares, respectively (column (8) of \cref{tab:demo_differences_exposure_rate}).

Interestingly, for age, the coefficients flip between the cumulative and rate (column (8) of \cref{tab:demo_differences_cum_exposure}). Older users (65+) have more of their visits tracked overall than younger users (18–24), according to the cumulative measure (column (8), \cref{tab:demo_differences_cum_exposure}). But when we look at the share of visits tracked, older users (50+) are less exposed than younger users—by at least 6.6 percentage points ($p < .01$). The difference reflects differences in browsing patterns by age.%
\footnote{
    As with \cref{sec:demo_diff_mechanisms}, the demographic differences in the depth of tracking by organizations for education levels and age groups persist after correcting for multiple demographic tests (see Footnote~\ref{fn:bonferroni}).
}
These findings reinforce the theme in \cref{sec:demo_diff_mechanisms}, where nearly all users are tracked online, but the intensity and structure of that tracking vary systematically by demographic characteristics.
The depth of tracking by big organizations (e.g., Google, Microsoft, Facebook) reflects not just differences in online behavior but also deeper patterns of the digital gap.

\section{Discussion}
\label{sec:discussion}

By linking digital traces from a representative sample of American adults with domain-level tracking audits, this study estimates individuals' exposure to online tracking. It also identifies who collects this information and how much of a user's web activity they can observe. The analysis advances the literature on online privacy in several ways.

First, unlike prior research that largely focused on audits of the most visited or most prominent websites \citep{Englehardt2016, karaj2018whotracks, blacklight, niforatos2021prevalence, SanchezRola2018, SanchezRola2021, zheutlin2021data, zheutlin2022data}, this study leverages passively observed browsing data from a large, representative sample.
This allows for a more accurate estimate of actual user-level tracking exposure across the population \citep{Dambra2022}.
\citet{Dambra2022} take a foundational step toward user-centric measurement by combining antivirus telemetry with custom web crawls, finding that user-level exposure is more concentrated than that measured from the trackers' perspective.
Our study complements and extends this approach by linking domain-level tracking data---covering a wide range of tracking technologies---to observed browsing behavior from a representative panel of American adults with demographic data, allowing us to examine unequal exposure by demographics.

Second, the findings confirm that tracking on the web is nearly universal.
Virtually all users in the sample encountered ad trackers and third-party cookies, with a median exposure in the tens of thousands. Like \cite{Dambra2022}, we find that these encounters occur rapidly.
Most users were exposed to these trackers within the first 48 hours of the month-long observation period. We further show that more invasive technologies---such as session recording, keylogging, and canvas fingerprinting---appear less frequently but are still widespread, with over 40\% of users encountering each of them within the first two days.

Third, exposure is not evenly distributed across the population. Users with more formal education, for instance, tend to experience higher levels of tracking. However, much of this disparity is explained by differences in browsing intensity. When exposure is normalized by the number of visits, demographic differences attenuate substantially, suggesting that more educated users are tracked more in part because they are online more often.

Yet, not all disparities vanish after accounting for browsing volume. In particular, older users consistently exhibit higher exposure rates per visit. This suggests that differences in exposure are not solely driven by time spent online, but also by the types of websites visited and the trackers embedded within them.

Despite these patterns, demographics explain only a small share of the variation in tracking exposure. Across both cumulative and normalized measures, the explanatory power of demographic variables is limited, with R-squared values of less than 8 percent in all specifications.

Finally, we examine the concentration of tracking across organizations. Although users may encounter hundreds of trackers, exposure is highly concentrated. 
We identify the same top three tracking organizations as \cite{Dambra2022}, which analyzes the top tracking organizations by aggregating over all visits.
As with \cite{Dambra2022}, we find Google the most pervasive.
Our estimates indicate that Google alone captures the largest share of browsing history for nearly 90\% of users, with a median share of 54\% of visits.
The next closest organizations---Microsoft and Facebook---are the dominant trackers for only about 4\% of users each, underscoring the extent to which a few firms dominate the tracking ecosystem.
Our analysis of organization tracking aggregates across users, identifying the single organization that observes the largest share of their browsing activity.
This user-level measure of organizational dominance further allows us to examine how concentration varies across demographic groups, revealing, for instance, that younger users have a higher proportion of their browsing history visible to a single organization.

Several limitations of the study warrant discussion. First, while the digital traces include activity from mobile phones, they do not cover the tracking ecosystems within mobile applications, which often rely on embedded software development kits (SDKs) not detectable via browser-based methods \citep{achara2015unicity, Binns2018}.

Second, the tracking audit tool, Blacklight, analyzes domains in real-time but has important blind spots. It does not detect more obfuscated forms of tracking, such as CNAME cloaking, nor does it capture server-side tracking that occurs outside the browser—even when users block cookies. Moreover, Blacklight focuses exclusively on client-side methods and may miss less visible forms of tracking. It also does not differentiate between benign and potentially harmful tracking; for example, session recording or canvas fingerprinting may be used for bot detection or UX testing, not necessarily surveillance \citep{blacklight, leakyforms}.

Third, tracking audits were successful for only about half of the visited domains. These successfully scanned domains account for more than 75\% of total visits, suggesting that failed scans occurred on less-visited sites. Additionally, a small subset of participants had no recorded web activity during the study period and were excluded from the analysis. In both cases, we assume that the missingness is unrelated to tracking exposure. While the high coverage of visits and low participant attrition reduce this concern, the possibility remains that tracking patterns differ systematically in the unobserved cases.

Fourth, the data rely on passive metering, and users' awareness of being observed—despite consenting to monitoring—may suppress true behavior. This could lead to an underestimation of actual tracking exposure, making our estimates conservative lower bounds \citep{Bosch2024, penney2016chilling, bad_domains, ss-adult}.

Finally, our exposure measures reflect potential visibility to third-party organizations, not confirmed data transfers or behavioral profiling, though the presence of trackers is widely used as a proxy for privacy risk \citep{Dambra2022, karaj2018whotracks, blacklight, niforatos2021prevalence, zheutlin2021data, zheutlin2022data}.

\section*{Ethics Statement}
The study protocol (CSU IRB Protocol 6404) was reviewed by the Colorado State University Institutional Review Board.

\section*{Code and Data}
To help others build on our work and to facilitate replication, we have released the code and data under an open source license at: \url{https://github.com/themains/private_blacklight}.

\FloatBarrier
\singlespacing
\clearpage
\bibliographystyle{apsr}
\bibliography{blacklight.bib}
\addcontentsline{toc}{section}{References}
\clearpage

\appendix
\onehalfspacing
\renewcommand{\thesection}{\Alph{section}}
\renewcommand\thetable{\thesection.\arabic{table}}
\renewcommand\thefigure{\thesection.\arabic{figure}}
\renewcommand\theequation{\thesection.\arabic{equation}}
\counterwithin{figure}{section}
\counterwithin{table}{section}
\counterwithin{equation}{section}

\begin{center}
\Large{Supporting Information}
\end{center}

\FloatBarrier
\section{Participant consent and data privacy}
\label{sm:yougov}
Before enrolling in a YouGov panel, people receive detailed information about the nature and scope of the data collection. Potential participants are informed about the types of data that will be collected, such as visited domains, and what will not be collected, including any information entered into secure forms, such as usernames, passwords, or payment details (See YouGov's FAQ, \url{https://today.yougov.com/about/faq}).

Only after reviewing this information do individuals consent to participate. Participation is entirely voluntary, and panelists can pause or uninstall the tracking software at any time. (See pages 3 and 4 of the \href{https://web.archive.org/web/20240530184843/https://d25d2506sfb94s.cloudfront.net/r/106/YouGovPulse_Windows_Installation_Manual.pdf}{installation guide} for Terms and Conditions and Privacy Policy made known to participants.) 

The browsing data collection application—YouGov Pulse---is developed in partnership with RealityMine and is available as a browser extension or mobile app. The app ensures anonymity: researchers never have access to identifying information, and no data is shared with third parties.

To encourage participation, panelists earn points through YouGov’s reward system--2,000 points upon joining and an additional 1,000 points for completing a full month of activity.

\includepdf[pages=-,scale=0.75, frame=true]{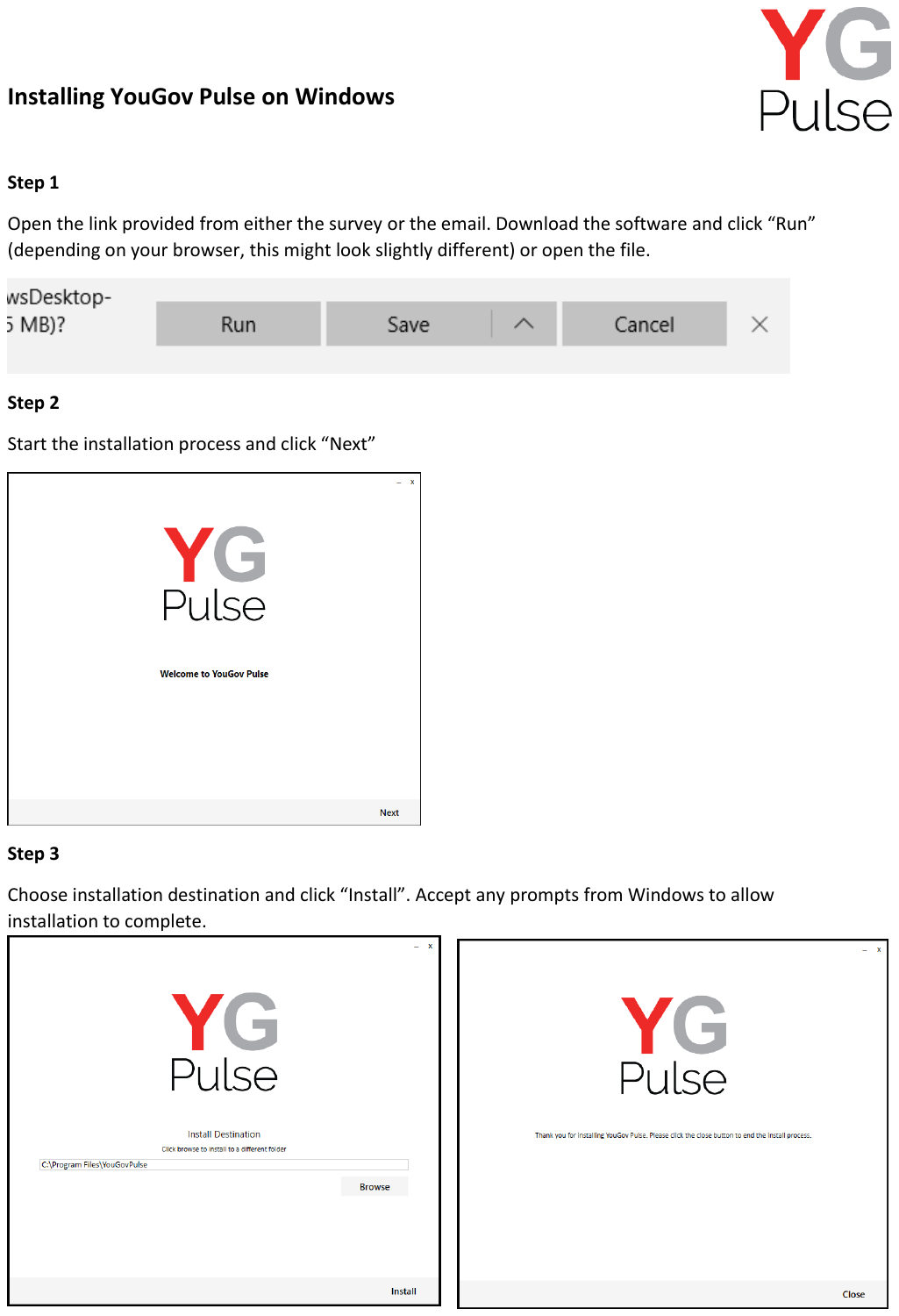}

\clearpage
\FloatBarrier
\section{Tracking methods}\label{sm:bl_tracking_mechanisms}
This appendix summarizes the seven tracking methods that Blacklight detects on the homepage of the domain and one additional randomly selected internal page \citep{blacklight}. 
Blacklight analyses are retrieved from a 24--48-hour cache when available or performed in real-time if no recent results exist.

\begin{itemize}
    \item{
        \textbf{Ad Tracking:} 
        Ad trackers are third-party scripts embedded in websites that collect user browsing behavior and send it to advertising networks. These scripts help build user profiles for targeted advertising or retargeting across websites. Blacklight detects ad tracking by identifying network requests to known advertising domains (domains under ``Ad Motivated Tracking'', \url{https://github.com/duckduckgo/tracker-radar/blob/main/docs/CATEGORIES.md}) in the DuckDuckGo Tracker Radar list.
    }
    \item{
        \textbf{Third-party Cookies:} 
        Cookies are small text files stored in the user's browser.
        Third-party cookies originate from domains other than the one being visited and are widely used to track users across websites.
    }
    \item{
        \textbf{Facebook Pixel:} 
        Facebook Pixel is a tracking script that monitors user behavior--such as page views, button clicks, and purchases---and sends this data to Facebook for ad targeting and conversion analytics.
        It links off-site behavior to user profiles across the Facebook ecosystem, even if users are not logged in to Facebook.
        Blacklight detects Facebook Pixel by identifying network requests to Facebook domains and inspecting URL query parameters for data patterns that match Pixel's documented schema.    
    }
    \item{
        \textbf{Google Analytics:} 
        Another major tracking tool operated by a major tech company is Google Analytics, which uses JavaScript tags and cookies to monitor user behavior such as session duration, navigation, and referrals.
        Blacklight detects it by flagging requests to known Google Analytics endpoints, such as \texttt{http://stats.g.doubleclick.net}.
    }

    \item{
        \textbf{Session Recording:} 
        Session replay scripts record user activity on a website, including mouse movements, scrolling, and form inputs---often in real time \citep{leakyforms}.
        These recordings can be replayed by website owners, revealing detailed behavioral data and potentially sensitive information.
        Blacklight detects session recording by monitoring network requests for URL substrings known to be associated with session replay tools (\url{https://web.archive.org/web/20210830151649/https://gist.github.com/gunesacar/0c67b94ad415841cf3be6761714147ca}).
    }
    \item{
        \textbf{Keylogging:} A potentially more invasive subset of session recording, keylogging captures every keystroke a user makes---including input into masked fields like passwords and credit card forms---before submission.
        This technique can reveal highly sensitive user data.
        Blacklight enters pre-determined text into input fields and monitors network requests for the same outgoing data.
    }
    \item{
        \textbf{Canvas Fingerprinting:}
        This method leverages the HTML5 canvas element to render invisible graphics and analyze subtle rendering differences based on the user's hardware and software configuration \citep{Acar2014, MS12}.
        These differences can be used to create a persistent, stateless identifier for tracking users across sessions \citep{karaj2018whotracks, blacklight}.
        Blacklight infers that canvas fingerprinting is used for tracking if scripts silently draw meaningful content on a sufficiently large canvas, do not use it for interactivity, and then extract pixel-level data in a way consistent with generating unique user identifiers.
    }
\end{itemize}

\clearpage
\FloatBarrier
\section{Alternative Visualization of Estimates}
\label{sm:coefplots}

\begin{figure}[ht]
  \centering
  \includegraphics[width=1.0\textwidth]{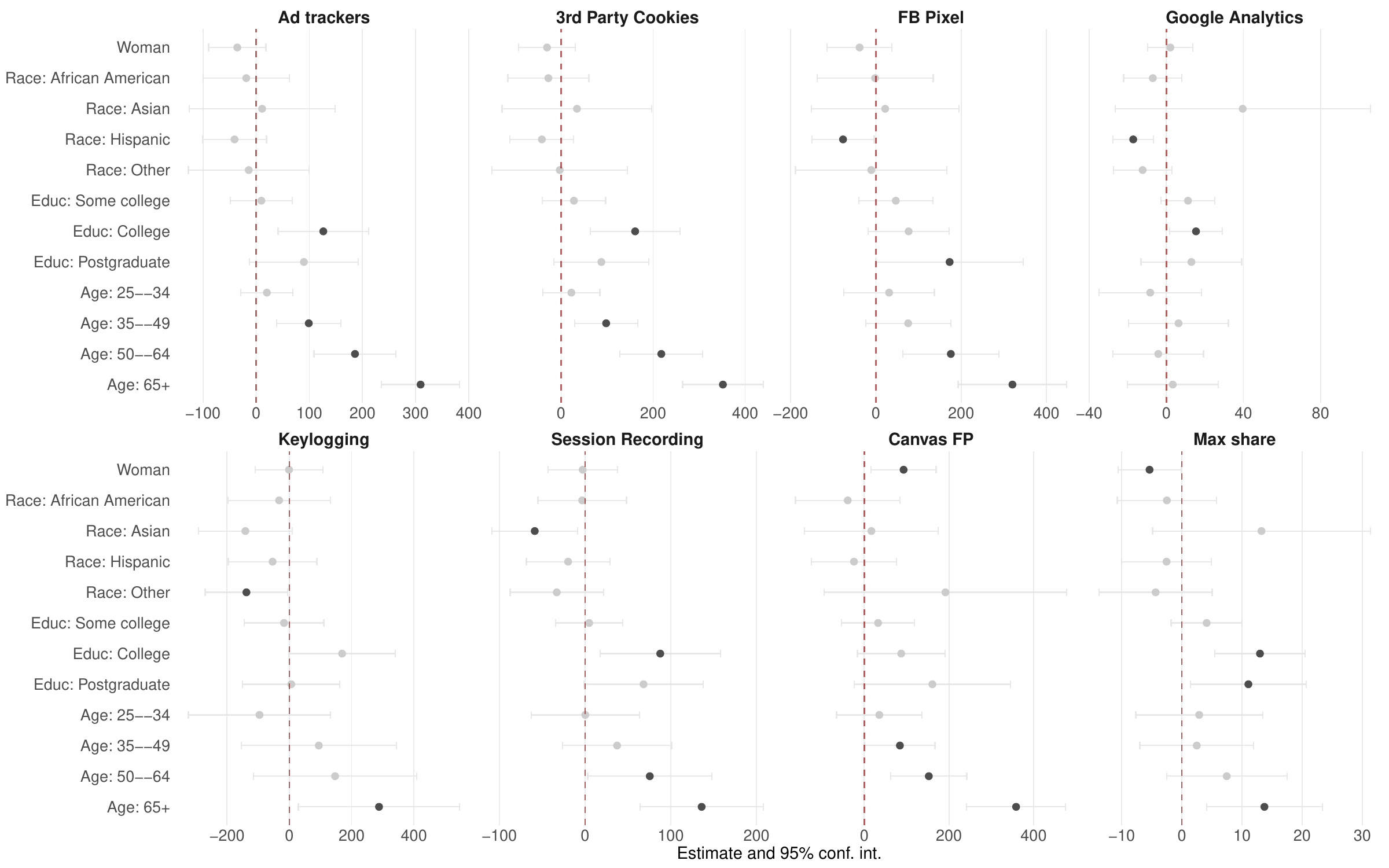}
  \caption{
    Estimated coefficients in \textit{cumulative exposure} by demographic group.
    Corresponds to \cref{tab:demo_differences_cum_exposure}.
    Each panel shows the estimated effect (makers) and 95\% confidence intervals (horizontal lines) from OLS regressions associating exposure to one of the seven tracking methods and the total number of visits tracked by a single organization.
    Black markers indicate statistically significant estimates at $p < .05$; gray markers indicate non-significant estimates.
  }
  \label{fig:coefplot_demo_differences_cumulative}
\end{figure}

\begin{figure}[ht]
  \centering
  \includegraphics[width=1.0\textwidth]{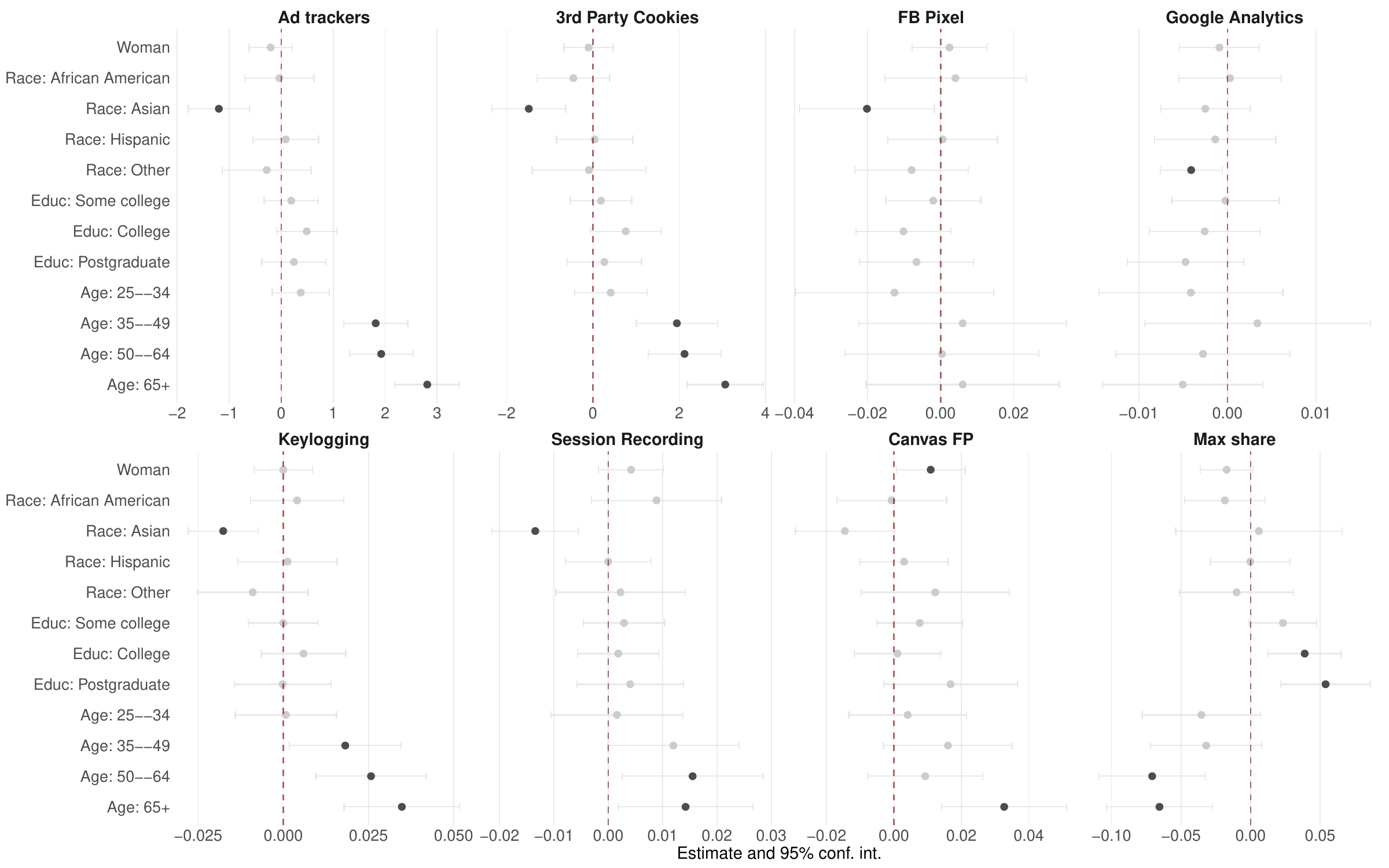}
  \caption{
    Estimated coefficients in \textit{exposure rate} by demographic group.
    Corresponds to \cref{tab:demo_differences_exposure_rate}.
    Each panel displays the OLS estimate and 95\% confidence intervals for regressions of either the exposure rate to the tracking technology or the proportion of visits tracked by a single organization. Black markers indicate statistically significant estimates at $p < .05$; gray markers indicate non-significant estimates.
  }
  \label{fig:coefplot_demo_differences_rate}
\end{figure}

\clearpage
\FloatBarrier
\section{Organization tracking weighted by time}
\label{sm:time_weighted_org_tracking}
\begin{figure}[ht]
  \centering
  \includegraphics[width=.55\textwidth]{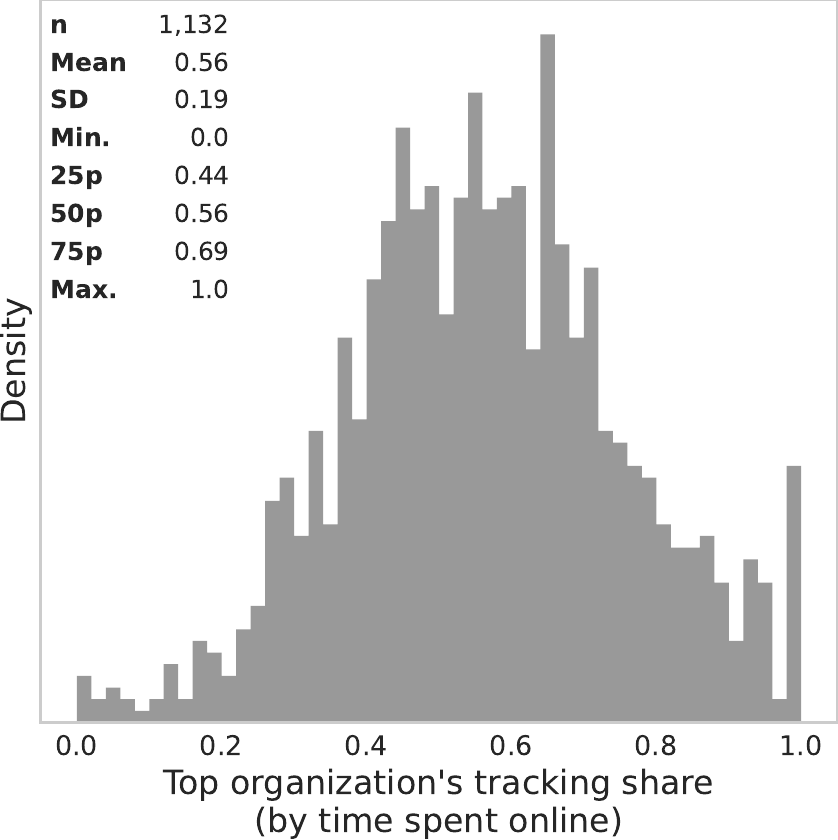}
  \caption{
    \small
    The largest share of each user's browsing time online tracked by a single organization (\cref{eq:ind_org_share_duration}):
    \begin{equation*}
            \text{Tracking share}_{ij}^{\text{(dur)}} = 
            \frac{
                \sum_{v \in \mathcal{V}_i} 
                \mathbf{1}
                (j \in O_{iv})
                \cdot
                t_{iv}
            }
            {
                \sum_{v \in \mathcal{V}_i} 
                t_{iv}
            }.
        \end{equation*}
    See \cref{fig:dist_maxshare_hist_summtable} for the corresponding figure for tracking shares by site visits.
  }
  \label{fig:dist_maxshare_hist_duration_summtable}
\end{figure}

\clearpage
\FloatBarrier
\section{Top Tracking Domains}
\label{sm:top_tracking_domains}
\newcolumntype{Y}{>{\raggedright\arraybackslash}X}
\begin{table}[ht]
\centering
\setlength\tabcolsep{2 pt}
\setlength{\defaultaddspace}{0pt}
\def\sym#1{\ifmmode^{#1}\else\(^{#1}\)\fi}
\caption{Top domains contributing to exposure}
\label{tab:bl_top_contributors_domain}
\begingroup
\centering
\begin{adjustbox}{max width=\textwidth}
\begin{tabular}{@{\hspace{0\tabcolsep}}llllllll@{\hspace{0\tabcolsep}}}
      \toprule
      &\multicolumn{1}{c}{Ads}&\multicolumn{1}{c}{Cookies}&\multicolumn{1}{c}{FB Pixel}&\multicolumn{1}{c}{GA}&\multicolumn{1}{c}{Session rec}&\multicolumn{1}{c}{Keyloggers}&\multicolumn{1}{c}{Canvas FP}\\
      &\multicolumn{1}{c}{(1)}&\multicolumn{1}{c}{(2)}&\multicolumn{1}{c}{(3)}&\multicolumn{1}{c}{(4)}&\multicolumn{1}{c}{(5)}&\multicolumn{1}{c}{(6)}&\multicolumn{1}{c}{(7)}\\
      \midrule
      1 & yahoo.com (246k) & yahoo.com (246k) & ebay.com (30k) & kohls.com (2.7k) & xfinity.com (10k) & yahoo.com (246k) & live.com (80k) \\
2 & google.com (987k) & google.com (987k) & capitaloneshopping.com (23k) & force.com (2.1k) & capitalone.com (9.9k) & capitaloneshopping.com (23k) & microsoft.com (26k) \\
3 & live.com (80k) & live.com (80k) & chase.com (14k) & pixiv.net (1.9k) & cbssports.com (6.2k) & smugmug.com (10k) & capitaloneshopping.com (23k) \\
4 & aol.com (47k) & bing.com (236k) & rakuten.com (12k) & mheducation.com (1.4k) & dell.com (5.5k) & weather.com (3.8k) & linkedin.com (19k) \\
5 & microsoft.com (26k) & microsoft.com (26k) & hulu.com (11k) & tupperware.com (1.4k) & att.com (4.9k) & activemeasure.com (3.6k) & rakuten.com (12k) \\
6 & cbssports.com (6.2k) & cbssports.com (6.2k) & xfinity.com (10k) & thriftbooks.com (1.0k) & earthlink.net (4.1k) & venatusmedia.com (3.5k) & hulu.com (11k) \\
7 & xfinity.com (10k) & xfinity.com (10k) & usps.com (9.7k) & adp.com (977) & venatusmedia.com (3.5k) & revenueuniverse.com (3.0k) & xfinity.com (10k) \\
8 & youtube.com (233k) & msn.com (39k) & nielseniq.com (9.4k) & equitybank.com (888) & homedepot.com (3.0k) & doceree.com (2.9k) & tiktok.com (10.0k) \\
9 & ebay.com (30k) & ebay.com (30k) & netflix.com (7.0k) & priceline.com (808) & doceree.com (2.9k) & spot.im (2.9k) & capitalone.com (9.9k) \\
10 & imdb.com (7.5k) & weather.com (3.8k) & wellsfargo.com (6.8k) & webtoons.com (705) & kohls.com (2.7k) & yelp.com (2.3k) & washingtonpost.com (8.0k) \\
11 & washingtonpost.com (8.0k) & dynata.com (22k) & dell.com (5.5k) & ourfamilywizard.com (614) & ancestry.com (2.6k) & attn.tv (2.2k) & espn.com (6.7k) \\
12 & rakuten.com (12k) & imdb.com (7.5k) & nextdoor.com (5.1k) & coupons.com (597) & discover.com (2.5k) & westlaw.com (2.1k) & target.com (5.9k) \\
13 & cnn.com (4.4k) & nielseniq.com (9.4k) & iheart.com (5.1k) & yaysavings.com (574) & zoosk.com (2.4k) & kroger.com (2.0k) & bankofamerica.com (5.7k) \\
14 & weather.com (3.8k) & cnn.com (4.4k) & 9gag.com (4.8k) & meetup.com (570) & attn.tv (2.2k) & ex.co (1.8k) & dell.com (5.5k) \\
15 & usps.com (9.7k) & youtube.com (233k) & earthlink.net (4.1k) & narvar.com (560) & cmix.com (2.0k) & dropbox.com (1.8k) & biggerbooks.com (4.0k) \\
16 & 9gag.com (4.8k) & twitter.com (111k) & biggerbooks.com (4.0k) & overdrive.com (557) & prizerebel.com (2.0k) & pnc.com (1.5k) & citi.com (3.9k) \\
17 & nielseniq.com (9.4k) & nytimes.com (6.0k) & activemeasure.com (3.6k) & managebuilding.com (529) & zleague.gg (1.9k) & morningjournal.com (1.1k) & cbsi.com (3.3k) \\
18 & nytimes.com (6.0k) & centurylink.net (1.8k) & venatusmedia.com (3.5k) & wootric.com (511) & trendmicro.com (1.9k) & 53.com (1.0k) & homedepot.com (3.0k) \\
19 & hulu.com (11k) & kohls.com (2.7k) & productreportcard.com (3.4k) & evergage.com (479) & phoenix.edu (1.9k) & thriftbooks.com (1.0k) & samsclub.com (2.8k) \\
20 & iheart.com (5.1k) & civicscience.com (7.4k) & cbsi.com (3.3k) & udemy.com (474) & verizon.com (1.8k) & trulia.com (995) & zulily.com (2.8k) \\
21 & kohls.com (2.7k) & dell.com (5.5k) & ups.com (3.2k) & fox.com (339) & jcpenney.com (1.5k) & qvc.com (991) & kohls.com (2.7k) \\
22 & foxnews.com (3.5k) & foxnews.com (3.5k) & homedepot.com (3.0k) & hobbylobby.com (311) & tupperware.com (1.4k) & dynatrace.com (922) & discover.com (2.5k) \\
23 & capitaloneshopping.com (23k) & aol.com (47k) & honeygain.com (2.9k) & daisous.com (309) & wurflcloud.com (1.4k) & newspapers.com (892) & adobe.com (2.4k) \\
24 & chase.com (14k) & rakuten.com (12k) & spot.im (2.9k) & wgal.com (308) & playsugarhouse.com (1.2k) & kaiserpermanente.org (890) & kroger.com (2.0k) \\
25 & dell.com (5.5k) & 9gag.com (4.8k) & samsclub.com (2.8k) & epsilon.com (292) & copart.com (1.1k) & mapquest.com (819) & shein.com (2.0k) \\
26 & dynata.com (22k) & google.co.uk (18k) & airbnb.com (2.8k) & noom.com (284) & veritonic.com (1.1k) & upmc.com (769) & trendmicro.com (1.9k) \\
27 & centurylink.net (1.8k) & chase.com (14k) & kohls.com (2.7k) & bizpacreview.com (274) & dominos.com (1.0k) & e-rewards.com (764) & aliexpress.com (1.8k) \\
28 & espn.com (6.7k) & capitalone.com (9.9k) & ancestry.com (2.6k) & factor75.com (245) & emi-rs.com (1.0k) & offerup.com (726) & pnc.com (1.5k) \\
29 & msn.com (39k) & morningjournal.com (1.1k) & discover.com (2.5k) & tbdliquids.com (234) & slickdeals.net (998) & odysee.com (700) & jcpenney.com (1.5k) \\
30 & linkedin.com (19k) & linkedin.com (19k) & adobe.com (2.4k) & reverbnation.com (224) & fidelity.com (975) & vccs.edu (653) & navyfederal.org (1.4k) \\
31 & twitter.com (111k) & nascar.com (903) & zoosk.com (2.4k) & avant.com (223) & newspapers.com (892) & forter.com (571) & coursera.org (1.4k) \\
32 & democraticunderground.com (14k) & spot.im (2.9k) & duolingo.com (2.4k) & mtsac.edu (218) & kaiserpermanente.org (890) & meetup.com (570) & ea.com (1.4k) \\
33 & zillow.com (19k) & investing.com (839) & vidyard.com (2.3k) & njlottery.com (211) & etrade.com (889) & blueconic.net (418) & nordstrom.com (1.3k) \\
34 & navyfederal.org (1.4k) & adobe.com (2.4k) & experian.com (2.2k) & examfx.com (198) & equitybank.com (888) & sutherlandglobal.com (403) & newyorklife.com (1.3k) \\
35 & oregonlive.com (1.4k) & zoho.com (15k) & attn.tv (2.2k) & gerberlife.com (197) & grabpoints.com (882) & reserveohio.com (399) & hp.com (1.2k) \\
36 & hideout.co (11k) & venatusmedia.com (3.5k) & kroger.com (2.0k) & higherincomejobs.com (193) & bhg.com (841) & bandcamp.com (375) & pusherapp.com (1.2k) \\
37 & zoosk.com (2.4k) & navyfederal.org (1.4k) & prizerebel.com (2.0k) & clover.com (182) & investing.com (839) & netspend.com (361) & playsugarhouse.com (1.2k) \\
38 & civicscience.com (7.4k) & huffpost.com (1.1k) & zleague.gg (1.9k) & onlygreatjobs.com (178) & gofundme.com (839) & freefarmtowngiftshop.com (339) & booking.com (1.1k) \\
39 & huffpost.com (1.1k) & trendmicro.com (1.9k) & trendmicro.com (1.9k) & kmov.com (177) & mcafee.com (817) & connatix.com (329) & expedia.com (1.1k) \\
40 & kitco.com (2.0k) & paycor.com (1.6k) & verizon.com (1.8k) & pushwoosh.com (172) & medallia.com (813) & hibid.com (329) & copart.com (1.1k) \\
41 & adobe.com (2.4k) & iheart.com (5.1k) & ex.co (1.8k) & truegloryhair.com (164) & adidas.com (783) & hobbylobby.com (311) & truist.com (1.0k) \\
42 & google.co.uk (18k) & cbsnews.com (865) & grizly.com (1.6k) & mintmobile.com (158) & chegg.com (767) & opentable.com (306) & 53.com (1.0k) \\
43 & capitalone.com (9.9k) & office.com (18k) & paycor.com (1.6k) & quantilope.com (157) & opera.com (757) & twinspires.com (306) & slickdeals.net (998) \\
44 & foodnetwork.com (1.0k) & attn.tv (2.2k) & allrecipes.com (1.6k) & walmart.com.mx (155) & wishpond.com (740) & ms.gov (296) & qvc.com (991) \\
45 & reddit.com (61k) & vidyard.com (2.3k) & westernjournal.com (1.5k) & yummybazaar.com (153) & neu.edu (724) & partycentersoftware.com (294) & adp.com (977) \\
46 & nascar.com (903) & bonvoyaged.com (751) & pnc.com (1.5k) & foxsports.com (148) & salemove.com (710) & pinnbank.com (259) & fidelity.com (975) \\
47 & morningjournal.com (1.1k) & doceree.com (2.9k) & mheducation.com (1.4k) & everyplate.com (146) & adam4adamsfw.com (697) & eyebuydirect.com (241) & citibankonline.com (971) \\
48 & ups.com (3.2k) & verizon.com (1.8k) & pandora.com (1.4k) & uscellular.com (144) & vergic.com (694) & centercode.com (236) & barclaycardus.com (928) \\
49 & discover.com (2.5k) & wellsfargo.com (6.8k) & oregonlive.com (1.4k) & pubnub.com (135) & pearson.com (672) & edx.org (216) & npr.org (922) \\
50 & westernjournal.com (1.5k) & meetup.com (570) & wurflcloud.com (1.4k) & guard.io (129) & oldnational.com (670) & chicoryapp.com (201) & michaels.com (900) \\
      \bottomrule
   \end{tabular}
\end{adjustbox}
\par\endgroup
\caption*{
    \footnotesize \textit{Note:}
    This table reports the top 50 domains (rows) contributing to individual-level exposure for each of the seven tracking methods (columns).
    A domain $d$'s contribution to individual-level exposure is computed as:
    \begin{equation*}
    \text{Contribution}_d^{(s)} = 
    \sum_{i} \sum_{v \in \mathcal{V}_{id}} 
    \vert trackers^{(s)}_d \vert,
    \end{equation*}
    based on all individual-domain visit instances, weighted by the number of trackers of type $s$ present on domain $d$.
    Parentheses report the total number of visits.
}
\end{table}

\end{document}